\documentclass[showpacs,aps,prb,reprint,superscriptaddress,nofootinbib,longbibliography]{revtex4-2}
\pdfoutput=1
\usepackage[colorlinks=true, pdfstartview=FitV,
bookmarks=true, bookmarksnumbered=true, breaklinks]{hyperref}
\usepackage{graphicx,color}

\usepackage{amsmath,amssymb,bm,color,longtable,mathrsfs,slashed,comment,tikz,ulem}

\usepackage{hyperref}
\hypersetup{
    colorlinks,
    linkcolor={red!50!black},
    citecolor={blue!50!black},
    urlcolor={blue!80!black}
}

\usepackage{footmisc}

\usepackage{xcolor}
\usepackage{cleveref}  
\usepackage{orcidlink}
\crefname{section}{Sec.}{sections}
\Crefname{section}{Section}{Sections}
\crefname{appendix}{App.}{appendices}
\Crefname{appendix}{Appendix}{Appendices}
\crefname{figure}{Fig.}{Figs.}
\Crefname{figure}{Figure}{Figures}
\crefname{equation}{Eq.}{Eqs.}
\Crefname{equation}{Equation}{Equations}
\newcommand{\phys}{\ensuremath{{d}}}
\newcommand{\spacedim}{\ensuremath{{dim}}}
\newcommand{\N}{\ensuremath{{N}}}
\newcommand{\nint}{\ensuremath{{n_\mathrm{int}}}}
\newcommand{\ndec}{\ensuremath{{n_\mathrm{dec}}}}
\newcommand{\lstein}{\ensuremath{{n_\mathrm{s}}}}
\newcommand{\order}{\ensuremath{{\mathcal{O}}}}

\def\be#1{\begin{equation}#1\end{equation}} 

\def\beqnn#1{\begin{eqnarray}#1\end{eqnarray}}

\begin{document}

\title{
Initial tensor construction and dependence of the tensor renormalization group on initial tensors
}

\author{Katsumasa~Nakayama\orcidlink{0000-0003-0270-8523}}
\email[]{katsumasa.nakayama@riken.jp}
\affiliation{RIKEN Center for Computational Science, Kobe, 650-0047, Japan}
\author{Manuel~Schneider\orcidlink{0000-0001-9348-8700}}
\affiliation{National Yang Ming Chiao Tung University (NYCU), Hsinchu, 30010, Taiwan}

\date{19 July 2024}

\begin{abstract}
We propose a method to construct a tensor network representation of partition functions without singular value decompositions nor series expansions.
The approach is demonstrated for one- and two-dimensional Ising models and we study the dependence of the tensor renormalization group (TRG) on the form of the initial tensors and their symmetries.
We further introduce variants of several tensor renormalization algorithms. Our benchmarks reveal a significant dependence of various TRG algorithms on the choice of initial tensors and their symmetries.
However, we show that the boundary TRG technique can eliminate the initial tensor dependence for all TRG methods. The numerical results of TRG calculations can thus be made significantly more robust with only a few changes in the code.
Furthermore, we study a three-dimensional $\mathbb{Z}_2$ gauge theory without gauge-fixing and confirm the applicability of the initial tensor construction.
Our method can straightforwardly be applied to systems with longer range and multi-site interactions, such as the next-nearest neighbor Ising model.

\end{abstract}


\maketitle

\tableofcontents

%
%
%
%
%
%
%
%
%
\section{Introduction}
Since its introduction about two decades ago~\cite{Levin:2006jai}, 
the tensor renormalization group (TRG) method was widely applied to statistical physics problems, including quantum field theories such as the CP(1) model~\cite{CP1TRG}, the $\mathbb{Z}_2$ gauge theory~\cite{Z2andmore,Kuramashi:2018mmi}, the Schwinger model~\cite{SchwingerOneFlavor,SchwingerTopological,SchwingerPhaseTransition}, and many more~\cite{Yu:2013sbi,Zou:2014rha,Yang:2015rra,GrossNeveu,Grassmann2018,Bazavov:2019qih,U1theta,Hirasawa2021,Akiyama2021,TRG_MultiFlavor,3dSU2chiral,SU2SU3}. 
The partition function is written in the form of a tensor network. Then, the partition function itself and other physical quantities can be calculated by contracting the tensor network, which means summing over all its indices. This is only possible using approximate methods which truncate the exponential growths of information with increasing the system size.
The tensor network is contracted in subsequent coarse-graining steps. In each step a truncation is applied, typically making use of a singular value decomposition (SVD). Since common algorithms coarse-grain the lattice in one direction only, the directions are exchanged after each step. The way this change is done affects the accuracy of the method and should therefore be carefully chosen. We discuss this effect in more details in \cref{app:xyrot}. Finally, physical quantities are extracted from the trace of the coarse-grained tensors.
Since the TRG is free of sampling problems~\cite{CP1TRG,SchwingerTopological,SchwingerPhaseTransition,Yang:2015rra,GrossNeveu,U1theta,Hirasawa2021,TRG_MultiFlavor,SchwingerTopological}, we can study systems for which Monte Carlo methods suffer from the sign problem~\cite{MonteCarloSignProblem}.

\paragraph{Overview of TRG algorithms.}

The TRG was originally introduced by Levin and Nave~\cite{Levin:2006jai}, and it was since improved by truncation methods that reduce the numerical costs~\cite{rand_trunc,Nakamura:2018enp,RandTRG,redsvd}.
The tensor network renormalization (TNR) additionally introduces disentanglers to improve the accuracy of the TRG~\cite{Evenbly2015}, an idea that originates in the multi-scale entanglement renormalization ansatz (MERA)~\cite{Jiang2008}.
For systems with relatively small volumes, the core TRG (CTRG) can also reduce the computational requirements~\cite{O4TRG}.

For higher dimensional systems, the TRG was extended to higher-order TRG (HOTRG)~\cite{HOTRG}.
Recently, various alternatives were also studied, such as the anisotropic TRG (ATRG)~\cite{ATRG}, the triad TRG (TTRG)~\cite{TriadRG}, and the minimally decomposed TRG (MDTRG)~\cite{MDTRG}. 
These methods can reduce the numerical costs, and allow for contractions of three- and higher-dimensional tensor network systems in feasible computational time. We explain several TRG algorithms in \cref{app:b_TRG,app:ATRG_and_Triad}

\paragraph{Initial tensor construction.}

In order to apply the TRG methods efficiently, we have to represent the physical quantities by a locally connected tensor network. This means that each index only appears on two neighboring tensors. 
Different geometries can arise for this network, depending on the connectivity of the interactions. We focus on square and cubic lattices. The TRG coarse-grains these lattices to a network with the same geometry and can thus be used iteratively.

Common approaches to construct a locally connected tensor network make use of SVDs or series expansions such as the Taylor expansion~\cite{Z2andmore,Baumgartner:2014nka,Marchis:2017oqi}. The expansion creates new variables, the power indices of each term. These can be used as indices of the initial tensor of the tensor network, by integrating out the original degrees of freedom. We give two examples of this construction in \cref{sec:InitialTensorDependence,sec:Z2}.

However, the choice of the initial tensors describing a given system is not unique. We propose another approach to construct the tensor network, based on a trivial decomposition with an identity matrix. The procedure does not require problem-specific and more involved decompositions, expansions, and variable transformations from spin indices to new tensor indices. We consider the spin indices as the indices of the initial tensor and localize the network by a matrix decomposition without approximations, inserting an identity matrix.
This method generally generates a local tensor network representation for a theory which can be described by a Lagrangian or Hamiltonian with periodic boundary conditions. However, the index dimension can be large, depending on the dimension of the local degrees of freedom and the range of the interaction. The method is very efficient for local interactions. The general resource scaling is discussed in \cref{app:general}.

\paragraph{Initial tensor dependence of TRG methods.}

Since the coarse graining steps include local truncations, the accuracy of TRG algorithms possibly depends on the form of the initial tensors.
Although the tensor construction based on the expansion is widely and successfully used for different models, it might not be the optimal choice for a given system and contraction method. We benchmark the accuracy of different TRG algorithms for the two-dimensional Ising model. Our results show that HOTRG-like methods, which use isometries for the coarse-graining step, are highly dependent on the symmetry of the initial tensors. We find that this problem does not apply when isometries in the algorithms are replaced by so-called squeezers~\cite{ATRG}, an idea originating from the boundary TRG~\cite{boundaryHOTRG}. This is possible for any isometry based TRG algorithm. Thus, we suggest to make use of this method in order to remove the dependence on the form of the initial tensors. In this case, our simple construction of the initial tensors leads to the same accuracy as other, more involved or problem-specific techniques. \Cref{app:b_TRG} discusses the technical details of how to implement squeezers in coarse-graining algorithms.

\paragraph{$\mathbb{Z}_2$ gauge theory.}

As an example for higher dimensional systems, the $\mathbb{Z}_2$ gauge theory in three spatial dimensions was studied with HOTRG and TRG~\cite{Kuramashi:2018mmi}.
There, the tensor network representation based on a Taylor expansion was used with gauge-fixing. The critical temperature was calculated with high accuracy.
However, the representation in~\cite{Kuramashi:2018mmi} has two-different tensors. Because the SVD is an optimization of local tensors, a smaller unit cell could generally be preferable. We show the applicability of our tensor network construction to the $\mathbb{Z}_2$ gauge theory, where only one initial tensor appears in the network. We calculate the free energy, specific heat and critical temperature without gauge-fixing, and find good agreement with previous calculations.

\paragraph{Structure of this paper.}

This paper is organized as follows.
We introduce our initial tensor construction in \cref{sec:IsingModel} for the one-dimensional Ising model with next-nearest neighbor interaction (NNNI) as a simple example. We can reproduce the exact solution with our method.
After this, we apply the method to the two-dimensional Ising model and study the initial tensor dependence of the TRG and HOTRG in \cref{sec:InitialTensorDependence}.
The accuracy of the HOTRG depends on the symmetricity of the initial tensor.
In \cref{sec:Z2} we apply the initial tensor construction method without gauge-fixing to the $\mathbb{Z}_2$ gauge theory and calculate the free energy and specific heat.
\Cref{app:general} explains how the method can be applied to general models and how the index sizes of the initial tensors scale.
We conclude our study in \cref{sec:Conclusion}.

%
%
%
%
%
%
%
%
%
\section{One-dimensional Ising model with next-nearest neighbor interactions\label{sec:IsingModel}}
We first introduce our method for the one-dimensional Ising model with next-nearest neighbor interactions and periodic boundary conditions as a simple example.
The idea has similarities to the initial tensor construction for a particular Ising model on a triangular lattice in~\cite{TNrep1}. Our method allows for other interaction terms than local interactions, such as the next nearest interactions in the case considered here.
The Ising model with NNNI in one spatial dimension with $N$ sites can be described by the partition function
\be{
Z
=
\sum_{\sigma=\pm1}\prod_{x=1} ^N T_{\sigma_x,\sigma_{x+1},\sigma_{x+2}} ^\mathrm{(1d)}. \label{eq:Ising1dPartitionFunction}
}
The sum $\sum_{\sigma=\pm1}$ indicates a summation over all combinations of the spins at all sites.
The tensor $T^{\mathrm{(1d)}}$ can be constructed with the spin indices $\sigma_x$ at sites $x$ and depends on the inverse temperature $\beta$ and the coupling constants $g_1$ and $g_2$:
\be{
	T_{\sigma_x,\sigma_{x+1},\sigma_{x+2}} ^\mathrm{(1d)}
		\equiv
		e^{-\beta (
			g_1\sigma_x \sigma_{x+1}
			+
			g_2\sigma_x \sigma_{x+2}
		)}.
}
This formulation does not form a locally connected network: a spin index $\sigma_{x}$ at a given site $x$ occurs on three different tensors ($T_{\sigma_{x-2},\sigma_{x-1},\sigma_{x}} ^\mathrm{(1d)}T_{\sigma_{x-1},\sigma_{x},\sigma_{x+1}} ^\mathrm{(1d)}T_{\sigma_x,\sigma_{x+1},\sigma_{x+2}} ^\mathrm{(1d)}$) instead of only two neighboring tensors.
Therefore, the partition function in \cref{eq:Ising1dPartitionFunction} cannot be used in typical coarse-graining algorithms. We have to find an alternative initial tensor formulation with only locally connected tensors.
For example, we want to find initial tensors $T'^{\mathrm{(1d)}}$ which only depend on two neighboring indices for a one-dimensional system.
For this, we first decompose the tensor $T^{\mathrm{(1d)}}$ into $A$ and $B$ without approximation by introducing a new index $a$:

\be{
	T_{\sigma_x,\sigma_{x+1},\sigma_{x+2}} ^\mathrm{(1d)}
		=
		\sum_{a_{x+1}=\pm1}
		A_{\sigma_x,\sigma_{x+1}} ^{a_{x+1}}
		B_{\sigma_{x+2}} ^{a_{x+1}}.
  \label{eq:Ising_ABdecompoaition}
}
We can apply a SVD or other methods for this decomposition, as it was previously done for an Ising model with a magnetic field in~\cite{TNrep1}.
However, we can also easily construct a tensor of the above form by choosing $B$ to be the identity matrix:
\be{
	T_{\sigma_x,\sigma_{x+1},\sigma_{x+2}} ^\mathrm{(1d)}
		=
		\sum_{a_{x+1}=\pm1}
		T_{\sigma_x,\sigma_{x+1},a_{x+1}} ^\mathrm{(1d)}
		\delta_{\sigma_{x+2}} ^{a_{x+1}}.
}

We then define the localized tensor in terms of the tensors $A$ and $B$, where the indices of tensor $B$ are shifted by one lattice site compared to \cref{eq:Ising_ABdecompoaition}: 
\be{
	T _{\sigma_x,\sigma_{x+1},a_x,a_{x+1}} '^{\mathrm{(1d)}}
		\equiv
		A_{\sigma_x,\sigma_{x+1}} ^{a_{x+1}}
		B_{\sigma_{x+1}} ^{a_{x}}.
}
Exploiting the periodic boundary conditions of the system, the partition function can be rewritten as a locally connected tensor network consisting of these new tensors:
\be{
	Z
		=
		\sum_{\sigma=\pm1}\sum_{a=\pm1}\prod_{x=1} ^NT '^{\mathrm{(1d)}} _{\sigma_x,\sigma_{x+1},a_x,a_{x+1}}.
}

By defining the combined indices $[a\sigma] \equiv a\otimes \sigma$, we obtain a one-dimensional system with size-four indices:

\be{
    Z
        =
        \sum_{[a\sigma ] = 1} ^4\prod_{x=1} ^{N}T'^{\mathrm{(1d)}} _{[a\sigma]_x,[a\sigma]_{x+1}}.
}

Since $T'^{\mathrm{(1d)}}$ is a $4\times 4$ matrix, we can easily find its eigenvalues by exact diagonalization. This gives the exact solution of the one-dimensional Ising model with NNNI, which is known from previous studies~\cite{Exact_NNNI,NNNI_3}.

Different ways of constructing the locally connected tensor network lead to different tensors $A$ and $B$ and thus different $T'$.
In general, we can relate different tensor representations of the same system using a unitary matrix:
\be{
    T^\mathrm{(new)} _{xx'}
        \equiv
        \sum_{k,k'}
        U _{xk} T^{\mathrm{(1d)}} _{kk'} U^\dagger _{k'x'}.
}
Although the partition function is analytically not changed by this transformation, the numerical accuracy of the coarse-graining steps can depend on the form of $T'^{\mathrm{(1d)}}$. This will be confirmed and studied in more detail in \cref{sec:InitialTensorDependence}.

The presented approach can be straightforwardly extended to an interaction with $\nint{}$ distinct hopping interactions.
Since each hopping term introduces a new index that gets combined with the spin index, the matrix size of $T'^{\mathrm{(1d)}}$ grows as $\phys{}^{\nint{}}$, where $\phys{}$ is the dimension of the spin index. See \cref{app:general} for more details.

One-dimensional models with several interactions are studied in the context of frustrated systems and antiferromagnetism~\cite{NNNI_1,NNNI_3,Zigzag_Ising,pendagonal,one_dim_Ising,NNNI_4,NNNI_5,1dim_experiment,random_Ising,long_range,NNNI_6}, and the approach presented here could be useful as a simple candidate to construct a locally connected network. 
Also, two-dimensional systems are widely studied to understand the phenomena of spin statistical systems~\cite{2dim_heisen,2dim_heisen_2,J1J2,J1J2_2,J1J2HOTRG,J1J2delta}. Our initial tensor construction can be readily applied to these systems. We show the explicit form of the initial tensors for the two-dimensional $J_1-J_2$ and $J_1-J_3$ Ising models in \cref{app:J1J2,app:J1J3} respectively.
In addition, we discuss more general systems including higher dimensions and long-range interaction in \cref{app:general}.
In principle, our construction can be extended to any dimension, and to various kinds of interaction terms.

%
%
%
%
%
%
%
%
%
\section{Two-dimensional Ising model and initial tensor dependence of the TRG methods\label{sec:InitialTensorDependence}}
We use the two-dimensional Ising model with periodic boundary conditions in a volume of $N\times N$ as a testing ground for the initial tensor dependence of different TRG methods. The partition function is
\begin{align}
	Z
	=&
		\sum_{\sigma=\pm1}
		\prod_{x,y=1} ^{N}
		e^{\beta h\sigma_{x,y}}
		e^{\frac{\beta g}{2}\sigma_{x,y}(\sigma_{x+1,y}+\sigma_{x,y+1})}\\
	=&
		\sum_{\sigma=\pm1}
		\prod_{x,y=1} ^{N}
		K_{\sigma_{x,y},\sigma_{x+1,y},\sigma_{x,y+1}},\label{eq:Ising2d_nonlocal}
\end{align}
with the spin indices $\sigma_{x,y}$ at sites $\{x,y\}$, the coupling constant $g$ and the external field $h$. In our numerical studies, we set $g=1$, $h = 0$, and $\beta = \beta_c = \mathrm{ln}(1 + \sqrt{
2})$, which is the critical value~\cite{IsingOnsager,IsingReview}.

\paragraph{Initial tensor construction with shifted delta-functions.}

The representation by the tensor $K$ is not a two-dimensional locally connected tensor network where the same index would only occur on two neighboring tensors. Thus, this formulation can not directly be used for the numerical evaluation of the partition function through coarse-graining algorithms.
We can construct a suitable network by inserting a delta function,

\be{
	Z
		=
		\sum_{a=\pm1}
		\sum_{\sigma=\pm1}
		\prod_{x,y=1} ^{N}
		K^{(\mathrm{delta})} _{\sigma_{x,y},\sigma_{x+1,y},a_{x,y},a_{x,y+1}},\label{eq:TN2dim}
}
where
\be{
	K^{(\mathrm{delta})} _{\sigma_{x,y},\sigma_{x+1,y},a_{x,y},a_{x,y+1}} \equiv K_{\sigma_{x,y},\sigma_{x+1,y},a_{x,y+1}}  
	\delta_{\sigma_{x,y}} ^{a_{x,y}}. \label{eq:2d_ising_Kdelta}
}
Similarly, other matrix decompositions like SVD or QR could be used instead of inserting a delta function. 
Again, we made use of the periodic boundary conditions and obtained a locally connected tensor network.

\paragraph{Initial tensor construction based on Taylor expansion.}

Previously, a different form of the initial tensor with $h=0$ was derived as in~\cite{Z2andmore,TNrep1,HOTRG}.
We explain it in the following as a reference to compare our method. We consider the Taylor expansion of a two site interaction in the Boltzmann weight. Because the square of a spin variable is the identity, only two factors in the expansion arise, which can be rewritten as a matrix multiplication:
\begin{align}
    &
    e^{(\beta g/2)\sigma_n\sigma_{n+1}}\nonumber\\ 
    =&
    \mathrm{cosh}(\beta g/2) + \sigma_n\sigma_{n+1}\mathrm{sinh}(\beta g/2)\nonumber \\
    =&
    \sum_{l=0} ^1 \Bigg(
    \sigma_n ^l \sqrt{\mathrm{cosh}(\beta g/2)}^{1-l}\sqrt{\mathrm{sinh}(\beta g/2)}^{l}\nonumber \\
    &\times
    \sigma_{n+1} ^l \sqrt{\mathrm{cosh}(\beta g/2)}^{1-l}\sqrt{\mathrm{sinh}(\beta g/2)}^{l} \Bigg)\nonumber \\
    =&
    \sum_{l=0} ^1
    W_{\sigma_n,l}
    W_{\sigma_{n+1},l}.
    \label{eq:taylorExpansion}
\end{align}
The matrix $W$ is defined as
\be{
	W
		=
		\begin{pmatrix}
    		 \sqrt{\mathrm{cosh}(\beta g/2)} & {\sqrt{\mathrm{sinh}(\beta g/2)}} \\
    		 {\sqrt{\mathrm{cosh}(\beta g/2)}} & {-\sqrt{\mathrm{sinh}(\beta g/2)}}
        \end{pmatrix},
\label{eq:costFunction}
}
where the first row corresponds to $\sigma = -1$, and the second to $\sigma = +1$. We see that the exponential of the two-site interaction can be decomposed into two $W$ matrices, introducing a new index $l$. Including the interaction terms in the orthogonal spatial direction, we get the initial tensor
\begin{align}
	&
	K^{(\mathrm{exp})}_{l_{x,y},l_{x+1,y},m_{x,y},m_{x,y+1}}
	\nonumber\\
	&=
	\sum_{\alpha}
	W_{\alpha,l_{x,y}}
	W_{\alpha,l_{x+1,y}}
	W_{\alpha,m_{x,y}}
	W_{\alpha,m_{x,y+1}}.
 \label{eq:2d_ising_Kexp}
\end{align}
This tensor is symmetric under the permutation of any indices.

\paragraph{Initial tensor dependence of TRG algorithms.}

We test the dependence of the coarse-graining methods on the initial tensors by using $K^{(\mathrm{delta})}$, $K^{(\mathrm{exp})}$, and the symmetrized tensor $K^{(\mathrm{sym})}$. The latter is obtained from $K^{(\mathrm{delta})}$ by a gauge transformation on the tensor indices, in order to make the tensor nearly symmetric under the permutation of its indices. This symmetrization is explained in \cref{app:sym_TRG}.
Each SVD in the coarse-graining step is truncated to a maximum bond size $D$ to prevent the exponential growth of the index sizes.
We apply $\order(D^6)$ TRG~\cite{Levin:2006jai}, $\order(D^7)$ HOTRG~\cite{HOTRG}, $\order(D^5)$ ATRG~\cite{ATRG}, $\order(D^5)$ MDTRG without internal line oversampling~\cite{MDTRG}, and $\order(D^7)$ boundary TRG for HOTRG (b-HOTRG)~\cite{boundaryHOTRG} for a system size of $V = 2^{20}$.

In this section we discuss the initial tensor dependence of the TRG, HOTRG, and b-HOTRG. The details of the algorithms and further benchmarks for the other TRG methods can be found in \cref{app:ATRG_and_Triad}.

\begin{figure}[htbp]
\begin{center}
 \includegraphics[width=8cm, angle=0]{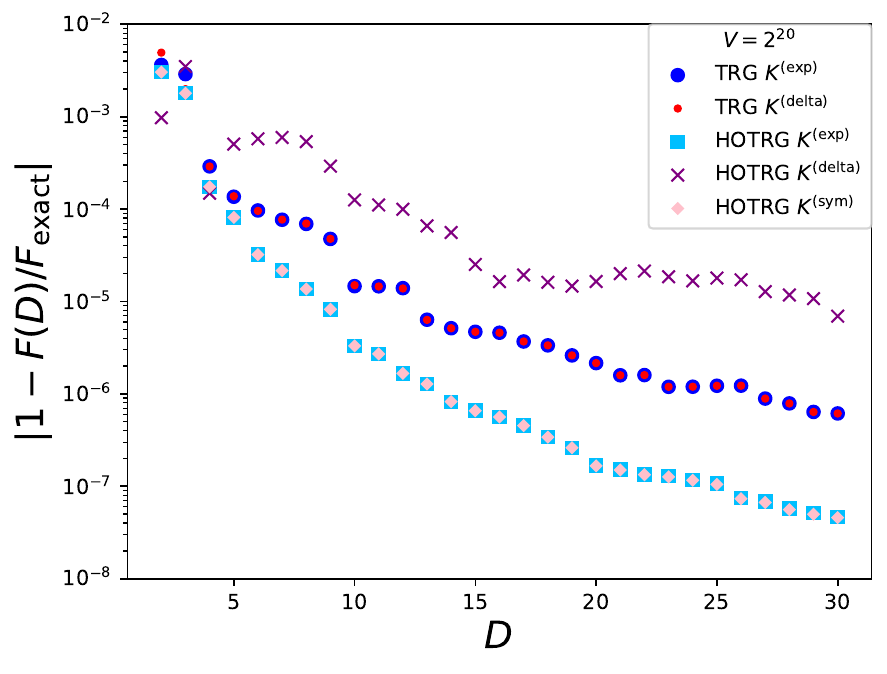}
 \caption{Dependence of the TRG and HOTRG methods on the form of the initial tensors for different cutoff bond dimensions $D$ in the two-dimensional Ising model. Shown are the relative errors of the free energy for the asymmetric initial tensor $K^{(\mathrm{delta})}$, the symmetric tensor $K^{(\mathrm{exp})}$, and the symmetrized tensor $K^\mathrm{(sym)}$. See main text for details.
}
\label{fig:TRG_HOTRG}
\end{center}
\end{figure}

\begin{figure}[htbp]
\begin{center}
 \includegraphics[width=8cm, angle=0]{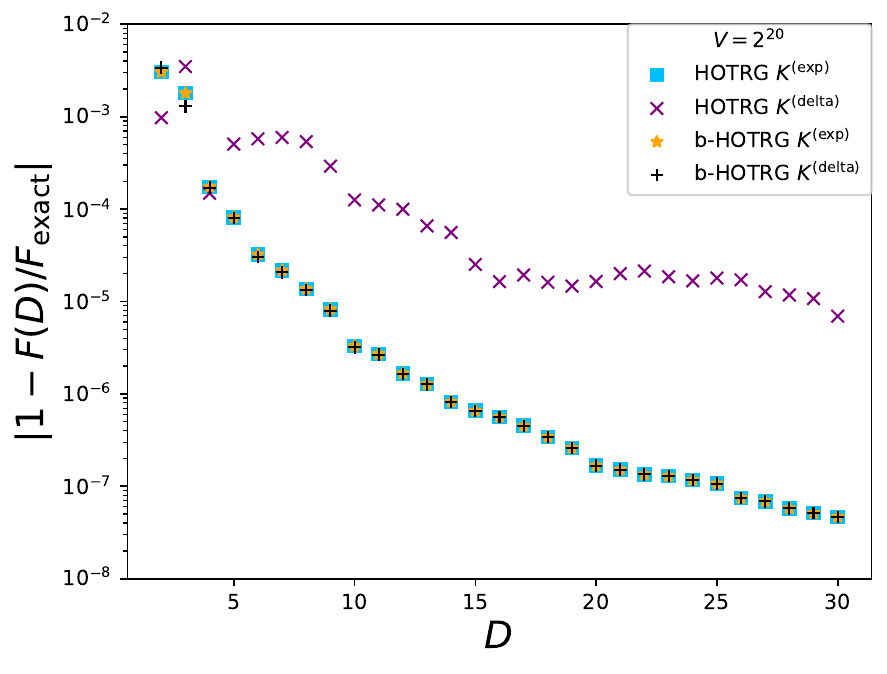}
 \caption{
 Dependence of the HOTRG and boundary-HOTRG methods on the form of the initial tensors for different cutoff bond dimensions $D$ in the two-dimensional Ising model. Shown are the relative errors of the free energy for the asymmetric initial tensor $K^{(\mathrm{delta})}$ and the symmetric tensor $K^{(\mathrm{exp})}$. The initial tensor dependence can be removed by introducing squeezers in the boundary HOTRG method. See main text and \cref{app:b_TRG} for details.
}
\label{fig:TRG_bHOTRG}
\end{center}
\end{figure}

We calculate the free energy $F\equiv-(\mathrm{ln}Z)/(\beta V)$ and compare it to the exact value~\cite{IsingExact}. 
\Cref{fig:TRG_HOTRG} shows the error of the free energy for the TRG and HOTRG methods.
We find that the accuracy of the original TRG method does not depend on the choice of the initial tensor. As in previous studies~\cite{HOTRG}, the HOTRG has better accuracy than TRG if the symmetric tensor $K^{(\mathrm{exp})}$ is used. The same holds for the symmetrized tensor $K^{(\mathrm{sym})}$. However, this is not true anymore for the asymmetric initial tensor $K^{(\mathrm{delta})}$, where the accuracy is lowered significantly. We study the symmetry dependence in more detail in \cref{app:sym_TRG} and find that the original TRG does generally not depend on the symmetry of the initial tensors, while HOTRG becomes more and more unreliable as less symmetric the initial tensors are.

\paragraph{Removing the initial tensor dependence by boundary TRG techniques.}

The HOTRG results for the asymmetric initial tensor can be improved by applying the boundary TRG method~\cite{boundaryHOTRG}. As shown in \cref{fig:TRG_bHOTRG}, this boundary HOTRG method produces results with the same accuracy as HOTRG for a symmetric initial tensor, but does so even if an asymmetric initial tensor $K^{(\mathrm{delta})}$ is used.
The boundary HOTRG differs from the simple HOTRG by the details of the coarse graining steps:
simple HOTRG uses an isometry $U^{(\mathrm{HOTRG})}$, while the boundary HOTRG introduces squeezers $P_1 ^{(\mathrm{bHOTRG})}$ and $P_2 ^{(\mathrm{bHOTRG})}$
for the coarse-graining. See \cref{app:b_TRG} for details.

Similar results are found for ATRG and MDTRG as shown in \cref{app:ATRG_and_Triad}.
We observe that ATRG and MDTRG with squeezers similar to the boundary TRG have no dependence on the form of the initial tensors, while coarse graining methods using isometries similar to the simple HOTRG strongly depend on it.

\paragraph{Overview of TRG methods and their initial tensor dependencies.}

We give a summary of the different coarse graining methods, their costs and their dependence on the initial tensors in \cref{tab:ATRGMDTRG}.
The different methods can be categorized into three classes. The first category uses no isometries but replaces tensors directly by their SVD representations, or by the projectors introduced in the boundary TRG method~\cite{boundaryHOTRG} (TRG, b-HOTRG). We call these projectors squeezers as in~\cite{ATRG} because they are not always projectors in the mathematical sense. We indicate this class of algorithms as \textit{sqz} in \cref{tab:ATRGMDTRG}.
The second category consists of methods that use the index of an isometry as a new index in the next coarse graining step (HOTRG-like). This is denoted as \textit{iso} in \cref{tab:ATRGMDTRG}. Finally, the third class consists of methods that use isometries for intermediate approximate contractions, but the indices of the isometries are not used as new indices of the coarse-grained tensors. We denote these methods as \textit{iso*}.

\begin{table}[h!]
 \centering
\begin{tabular}{l l l l l } 
 \hline \hline
   & Costs & Trun. & Dep. & $|1 - F(D=30)/F_\mathrm{ex}|$ \\ 
\hline
TRG~\cite{Levin:2006jai}   &$\order(D^6)$ & \textit{sqz} & $--$ & $\sim \order(10^{-6})$ \\
\hline
HOTRG~\cite{HOTRG}& $\order(D^{4\spacedim-1})$ & \textit{iso} & $++$ & $\order(10^{-5} \sim 10^{-8})$ \\
b-HOTRG~\cite{boundaryHOTRG} &  & \textit{sqz} & $--$ &$\sim \order(10^{-8})$  \\
\hline
ATRG~\cite{ATRG}& $\order(D^{2\spacedim+1})$ & \textit{sqz} & $--$ & $\sim \order(10^{-7})$\\
Iso-ATRG~\cite{ATRG}&  & \textit{iso} & $++$ & $\order(10^{-5} \sim 10^{-6})$\\
sh-ATRG &  & \textit{sqz} & $-$ &$\sim \order(10^{-7})$ \\
sh-Iso-ATRG &  & \textit{iso*} & $-$ & $\sim \order(10^{-6})$\\
\hline
MDTRG~\cite{MDTRG} & $\order(D^{\spacedim+3})$ & \textit{iso} & $++$ & $\order(10^{-5} \sim 10^{-7})$\\
sh-MDTRG &  & \textit{iso*} & $-$ & $\sim \order(10^{-6})$\\
b-MDTRG &  & \textit{sqz} & $--$ & $\sim \order(10^{-7})$\\
\hline \hline  
 \end{tabular}
 \caption{Properties of different TRG coarse graining methods. 2nd column: numerical costs; $D$ is the bond dimension and $\spacedim$ the spacetime-dimension. 3rd column: truncation method; \textit{iso} stands for isometries which are used to create the coarse-grained indices; \textit{iso*} means that isometries are used for intermediate approximate contractions, but they do not create the new indices of the coarse-grained tensors directly; \textit{sqz} denotes all other methods, so either the squeezers from boundary TRG~\cite{boundaryHOTRG} (see main text and \cref{app:b_TRG}), or a simple contraction and singular value decomposition. 4th column: dependence on the initial tensors; $--$ stands for no dependence, $-$ for a slight but not significant dependence, $++$ for strong dependence; 5th column: relative error for a bond dimension of $D=30$ for the two-dimensional critical Ising model compared to the exact energy; this gives an estimate of the accuracy, but note that different methods scale differently in the bond dimension. See \cref{app:ATRG_and_Triad} for more details on algorithms and benchmarks.
 } \label{tab:ATRGMDTRG}
\end{table}

From our calculations in \cref{fig:TRG_HOTRG,fig:TRG_bHOTRG,app:ATRG_and_Triad} we conclude that coarse graining methods making use of isometries to create the new indices (\textit{iso}), such as the simple HOTRG, depend strongly on their symmetry properties. This was also found for the massless Schwinger model with a different approach~\cite{PhysRevD.101.094509}. The isometries can always be replaced by squeezers as introduced for the boundary TRG. We suggest using these boundary TRG techniques, which can remove the dependence on the initial tensor symmetries and make the algorithm more robust (\textit{sqz}). In that case, our tensor construction provides a simple and generic way to represent the partition function as a locally connected tensor network, without loss of accuracy in numerical calculations compared to other construction techniques.

\paragraph{Dependence on the index exchange type.}
We also found a dependence of TRG algorithms on the way the index-directions are exchanged after each coarse graining step. From \cref{app:xyrot} we conclude that the exchange of directions should ideally allow the initial SVDs in a coarse graining step to split tensors along the contraction direction in the previous step. For the algorithms in this paper, this means that a rotation in clockwise or counterclockwise direction is better suited for shifted TRG methods. For non-shifted methods, a flip $x \leftrightarrow y$ ($x' \leftrightarrow y'$) leads to similar or better results. It replaces the $x$-index in negative (positive) $x$-direction with a corresponding $y$-index. We found that these flips lead to inaccurate results for the shifted methods and an accumulation of systematic errors. Therefore, the type of index exchange should be carefully checked for the TRG method used. In our benchmarks and numerical results we always apply the optimal exchange between directions, which is a rotation for shifted and a flip for non-shifted methods.

%
%
%
%
%
%
%
%
%
\section{$\mathbb{Z}_2$ gauge theory\label{sec:Z2}}

\begin{figure}[htbp]
    \begin{center}
         \includegraphics[width=3cm, angle=0]{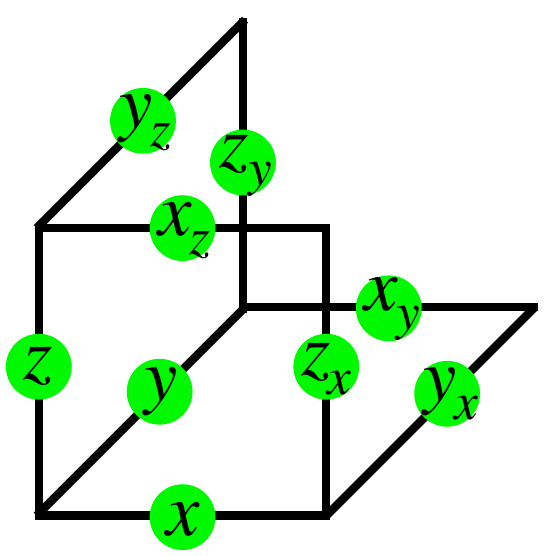}
         \caption{
        Schematic picture of the three plaquette terms at a lattice point $n$ in different planes for the three-dimensional $\mathbb{Z}_2$ gauge theory. Index notation as introduced in \cref{eq:Z2indexNotation}.
         }
        \label{fig:plaquette}
    \end{center}
\end{figure}

The three-dimensional $\mathbb{Z}_2$ gauge theory was studied in~\cite{Z2andmore,Kuramashi:2018mmi} using HOTRG and TRG. The partition function can be written as
\be{
    Z
        =
        2^{-3V}
        \sum_{\sigma=\pm1}
        \prod_{n,\mu > \nu}
        e^{
            -\beta
            \sigma_{n,\mu}
            \sigma_{n+\hat{\mu},\nu}
            \sigma_{n+\hat{\nu},\mu}
            \sigma_{n,\nu}
        },
}
where we introduce the link variables $\sigma_{n,\mu}$ at site $n$ with direction $\mu$. 
The unit vector in $\mu$ direction is represented by $\hat{\mu}$. The interaction corresponds to a spin system where each spin interacts with its nearest and next-nearest neighbors in a four-site interaction, known as plaquette-term. A schematic picture of the three plaquette terms in the three directions can be seen as black lines in \cref{fig:plaquette}.

\paragraph{Initial tensor construction based on Taylor expansion.}

In~\cite{Kuramashi:2018mmi}, the authors used a representation based on the Taylor expansion similar to \cref{eq:taylorExpansion}:
\begin{align}
    &e^{
        \beta
        \sigma_{n,\mu}
        \sigma_{n+\hat{\mu},\nu}
        \sigma_{n+\hat{\nu},\mu}
        \sigma_{n,\nu}
    }
    \nonumber\\
    &=
    \mathrm{cosh}\beta
    \sum_{p=0} ^1
        \left(\mathrm{tanh}\beta\right)^p
        \left(
            \sigma_{n,\mu}
            \sigma_{n+\hat{\mu},\nu}
            \sigma_{n+\hat{\nu},\mu}
            \sigma_{n,\nu}
        \right)^p.
\end{align}
In this previous work, a gauge-fixing was applied to simplify the tensor network representation. However, for gauge theories on the lattice in general, numerical calculations with gauge-fixing can suffer from the ambiguity of Gribov copies~\cite{Gribov:1977wm}. Therefore, we do not fix the gauge in our initial tensor constructions and in our numerical calculations.

Following the derivation in~\cite{Kuramashi:2018mmi,Z2andmore} but without gauge-fixing, we define the tensors $A$ and $B$ as

\be{
    A_{pqrs}
        =
        \mathrm{mod}(1+p+q+r+s,2)
}
\be{
    B_{pqrs}
        =
        (\mathrm{tanh}\beta)^{(p+q+r+s)/4}
        \delta_{pq}
        \delta_{qr}
        \delta_{rs}.
}
A combination of six tensors leads to a unit cell tensor $T^{(\mathrm{exp})}$ which defines a locally connected tensor network that reproduces the partition function:
\begin{align}
    &
    T^{(\mathrm{exp})} _{[xX][x'X'][yY][y'Y'][zZ][z'Z']}/(\mathrm{cosh}\beta)^3
    \equiv \nonumber\\
        &
        \sum_{a,b,c,d,e,f}
        A _{cyZe}
        A _{fzxb}
        A _{dYXa}
        B _{bx'y'c}
        B _{aX'Z'e}
        B _{fz'Y'd} .
    \label{eq:Z2Texp}
\end{align}
The combination of two indices like $[xX]\equiv x\otimes X$ introduces new spin-3/2 indices for the unit cell tensor.

Note that $T^{\mathrm{(exp)}}$ is not symmetric, even if $A$ and $B$ are completely symmetric in all indices.
This differs from the Ising model, where the initial tensors obtained using a Taylor expansion were symmetric. Therefore, the expansion method does not produce better symmetry properties than our method for the $\mathbb{Z}_2$ model. 
For the Ising model, we found in \cref{sec:InitialTensorDependence,app:ATRG_and_Triad} that HOTRG is not well suited for non-symmetric initial tensors, while ATRG does not depend on the symmetry properties. This suggests that ATRG is a better choice for the initial tensors $T^{\mathrm{(delta)}}$ and $T^{\mathrm{(exp)}}$of the $\mathbb{Z}_2$ model. However, the initial tensor dependence was not explicitly checked for the $\mathbb{Z}_2$ model and we use ATRG in all our simulations.

The previous study~\cite{Kuramashi:2018mmi} applied further constraints on the tensors A and B to implement a gauge-fixing condition. With this, they could precisely reproduce Monte-Carlo results.

\paragraph{Initial tensor construction with shifted delta-functions.}

In the following, we construct another tensor network for the same model using the method introduced in \cref{sec:IsingModel}. We do not need a Taylor expansion, do not make use of the spin property $\sigma ^2 = 1$, and keep the gauge unfixed.
For a simpler notation, we define the indices
\begin{align}
    x_{\hat{k}}
    \equiv&
        \sigma_{n+\hat{k},\mu=0}\nonumber\\
    y_{\hat{k}}
    \equiv&
        \sigma_{n+\hat{k},\mu=1}\\
    z_{\hat{k}}
    \equiv&
        \sigma_{n+\hat{k},\mu=2}.\nonumber
    \label{eq:Z2indexNotation}
\end{align}
 Moreover, we define $x \equiv x_{\hat{0}}$, $y \equiv y_{\hat{0}}$, $z \equiv z_{\hat{0}}$.
 The index $n$ is not written explicitly here for brevity.
\Cref{fig:plaquette} shows a graphical representation of this index convention, where we locate the degrees of freedom $\sigma_{n,\mu}$ on the links between sites $n$ and $n+\hat{\mu}$.

The Boltzmann weight at site $n$ is
\begin{align}
&T_{x,x_{\hat{y}},x_{\hat{z}},y,y_{\hat{x}},y_{\hat{z}},z,z_{\hat{x}},z_{\hat{y}}}
\nonumber\\
\equiv&
e^{
-\beta
\left(
x
x_{\hat{y}}
y
y_{\hat{x}}
+
x
x_{\hat{z}}
z
z_{\hat{x}}
+
y
y_{\hat{z}}
z
z_{\hat{y}}\right)
}/8 .
\end{align}

We can translate this weight to a tensor network.
For example, we can split the index $x_{\hat{y}}$ from the tensor:
\begin{align}
&T_{
x,x_{\hat{y}},x_{\hat{z}},
y,y_{\hat{x}},y_{\hat{z}},
z,z_{\hat{x}},z_{\hat{y}}
}
\nonumber\\
=&
\sum_{{a_{\hat{y}}} = \pm 1}
A_{
x,x_{\hat{z}},
y,y_{\hat{x}},y_{\hat{z}},
z,z_{\hat{x}},z_{\hat{y}}
} ^{a_{\hat{y}}}
B_{x_{\hat{y}}} ^{a_{\hat{y}}} 
\end{align}
One of the simplest choices for this decomposition is $B_{x_{\hat{y}}} ^{a_{\hat{y}}} = \delta_{a_{\hat{y}},x_{\hat{y}}}
$.
We define a new tensor without summation of the index $x$,
\begin{align}
&
C_{
x,x_{\hat{z}},a,a_{\hat{y}},
y,y_{\hat{x}},y_{\hat{z}},
z,z_{\hat{x}},z_{\hat{y}}
}
\nonumber\\
\equiv &
T_{
x,a_{\hat{y}},x_{\hat{z}},
y,y_{\hat{x}},y_{\hat{z}},
z,z_{\hat{x}},z_{\hat{y}}
}
\delta _{x} ^{a},
\end{align}
where $a \equiv a_{\hat{0}}$.
Similarly, we can split the indices $y_{\hat{z}}$ and $z_{\hat{x}}$ from the tensor and shift the indices. This way, we obtain the initial tensor
\begin{align}
& T ^{(\mathrm{delta})} _{
x,x_{\hat{z}},a,a_{\hat{y}},
y,y_{\hat{x}},b,b_{\hat{z}},
z,z_{\hat{y}},c,c_{\hat{x}}.
}
\nonumber\\
\equiv& 
T_{
x,x_{\hat{z}},a_{\hat{y}},
y,y_{\hat{x}},b_{\hat{z}},
z,z_{\hat{y}},c_{\hat{x}}
}
\delta _{x} ^{a}
\delta _{y} ^{b}
\delta _{z} ^{c}.
\label{eq:Z2Tdelta}
\end{align}
We define new spin-3/2 indices, $[az]_{n+\hat{y}} \equiv a_{\hat{y}}\otimes z_{\hat{y}}$ and finally obtain the partition function

\be{
Z
=
\sum_{[az],[bx],[cy] = 1} ^4
\prod_{n}
T  ^{(\mathrm{delta})} _{
[az]_{n}[az]_{n+\hat{y}}
[bx]_{n}[bx]_{n+\hat{z}}
[cy]_{n}[cy]_{n+\hat{x}}
}.
\label{new_rep}
}
This is a locally connected tensor network representation. 

\paragraph{Numerical results for the free energy.}

We test this representation with the initial tensor $T  ^{(\mathrm{delta})}$ without gauge-fixing by evaluating the partition function numerically.
We set the system sizes in $x$, $y$, $z$ direction to $N_x = 2$, $N_y = N_z = 2^{15}$. The first dimension is chosen small, similarly to~\cite{Kuramashi:2018mmi}.
First, the three-dimensional system is reduced to a two-dimensional one by an HOTRG step without truncation.
Then, we apply ATRG to perform the coarse-graining contractions with a truncation at a given bond dimension $D$.

\begin{figure}[htbp]
    \begin{center}
         \includegraphics[width=7cm, angle=0]{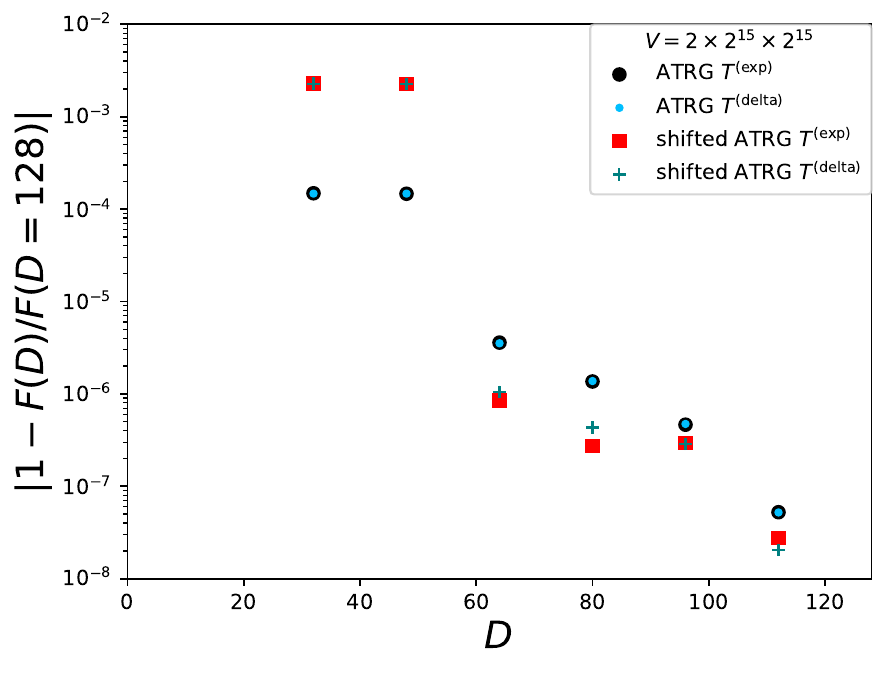}
         \caption{
         Relative error of the free energy $F$ for the three-dimensional $\mathbb{Z} _2$ gauge theory for different bond dimensions $D$. Shown are the results for ATRG and the shifted ATRG (see \cref{app:ATRG_and_Triad}) for the initial tensors $T^{(\mathrm{exp})}$ from \cref{eq:Z2Texp} and $T^{(\mathrm{delta})}$ from \cref{eq:Z2Tdelta}.
         }
        \label{fig:free_ene}
    \end{center}
\end{figure}

The free energy
\be{
    F \equiv
        -\frac{1}{\beta V}
        \mathrm{ln}Z,
}
is calculated from the partition function. The relative error in dependence on the cutoff parameter $D$ is estimated by $|1 - F(D)/F(D=128)|$, where $D=128$ is the largest bond dimension in our simulations.

\Cref{fig:free_ene} shows the error for the ATRG coarse graining method at $\beta = 0.6561$ with oversampling parameter $r = 2$. Additionally, we show the results for the shifted ATRG algorithm, which is explained in \cref{app:ATRG_and_Triad}.
We observe no significant dependence on the initial tensor for both methods.
The initial tensor $T^\mathrm{(delta)}$, which is constructed without a Taylor expansion, leads to accurate results and the accuracy is comparable to calculations with the initial tensor $T^\mathrm{(exp)}$.
The relative error between the ATRG and shifted ATRG methods is $\left| 1 - \frac{F_\mathrm{sh,ATRG}(D=128)}{F_\mathrm{ATRG}(D=128)} \right| = \order(10^{-7})$, indicating that both methods converge to the same value.
The error from randomized SVDs is sufficiently reduced by an $r=2D$ oversampling.
Since the shifted ATRG is better suited for the impurity tensor method, as discussed in \cref{app:imp_ATRG}, we use the shifted ATRG in calculations of the specific heat of the system.

The calculation of the free energy for the three-dimensional $\mathbb{Z}_2$ model demonstrates that our initial tensor construction $T^\mathrm{(delta)}$ without expansion and gauge-fixing leads to results as accurate as those with the initial tensor constructions $T^\mathrm{(exp)}$.

\begin{figure}[htbp]
    \begin{center}
        \includegraphics[width=7cm, angle=0]{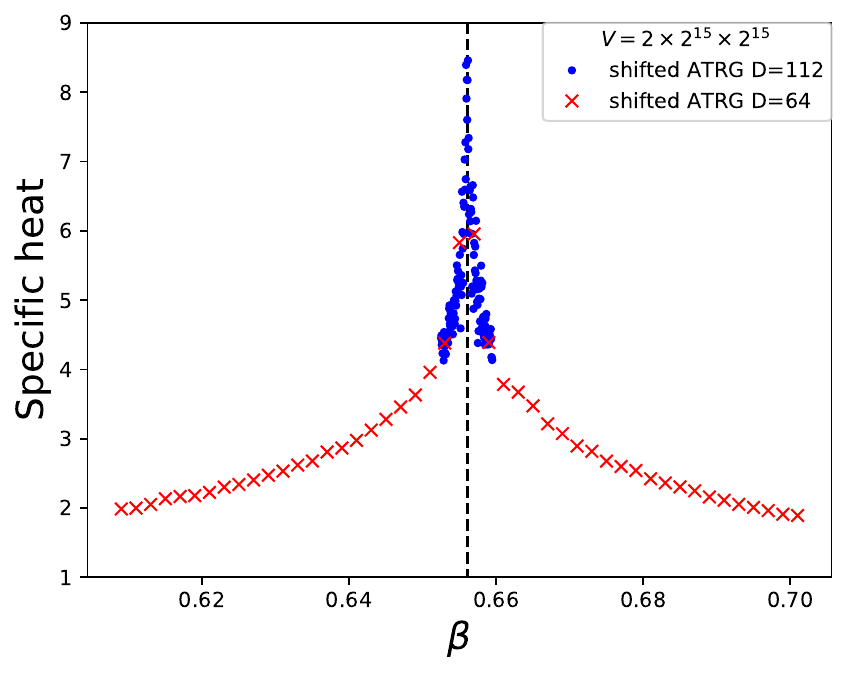}
        
        (a)
        
         \includegraphics[width=7cm, angle=0]{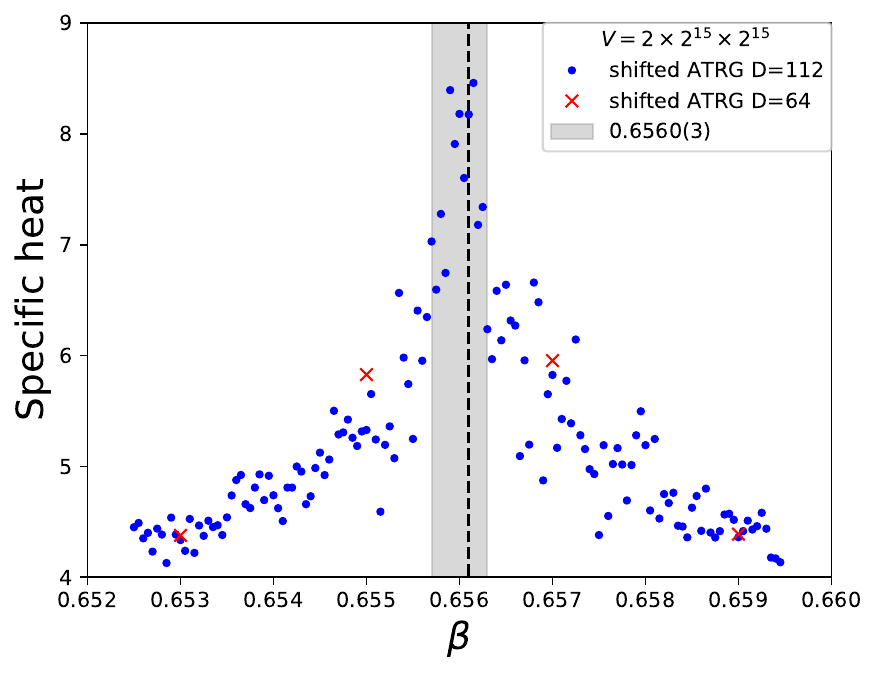}
         
         (b)
         
         \caption{
            Specific heat of the three-dimensional $\mathbb{Z} _2$ gauge theory for different inverse temperatures $\beta$. The dotted line marks the critical temperature $\beta_c = 0.6561$ calculated in~\cite{Kuramashi:2018mmi}. Shown are the results obtained with the shifted ATRG method (see \cref{app:ATRG_and_Triad}) and the initial tensors $T^{(\mathrm{delta})}$ from \cref{eq:Z2Tdelta}. See the main text for details of the calculation. (a) wide range of $\beta$. (b) zoom-in around the maximum value. Neighboring data points fluctuate due to the statistical errors from the randomized SVD. The grey band is taken as an estimate of the uncertainty of $\beta_c$. See main text for details.
         }
        \label{fig:spe_heat}
    \end{center}
\end{figure}

\paragraph{Numerical results for the specific heat.}

We further calculate the specific heat
\be{
    C
    \equiv
        \beta ^2
        \frac{1}{V}
        \frac{\partial^2 \mathrm{ln}Z}{\partial \beta ^2}.
}
First, we obtain the first order derivative $\partial_\beta\mathrm{ln}Z$ by the impurity tensor method as explained in \cref{app:imp_ATRG}. Then, the second order derivative and therefore $C$ is derived from this with a numerical forth-order approximation of the differentials. For calculations not too close to the critical temperature, we choose a step size of $\delta \beta = 0.002$ and a bond dimension $D=64$. Closer to the critical value of $\beta$ we set $\delta \beta = 0.00025$ and the bond dimension to $D=112$.
The error of the approximation for the second order derivative is $\order(\delta \beta ^4)$, becoming small for smaller $\delta \beta$. On the other hand, any kind of error of the first order derivative $\delta(\partial_\beta\mathrm{ln}Z)$ propagates as $\order(\delta (\partial_\beta\mathrm{ln}Z) /(\delta \beta))$, growing for small $\delta \beta$. If one aims for high precision, the step size $\delta \beta$ should therefore be carefully chosen and optimized.

\Cref{fig:spe_heat} shows the specific heat of the $\mathbb{Z}_2$ gauge theory with the initial tensor $T^\mathrm{(delta)}$.
The critical temperature is found to be $\beta_c = 0.6560(3)$.
The uncertainty is estimated by the spread of results due to the randomized SVD. We choose the uncertainty of $\beta_c$ such that the largest ten data points lie in the error band, see \cref{fig:free_ene}(b). A more careful study of error sources would be needed if one aims for higher precision. Further methods to improve the accuracy can be found in~\cite{Kuramashi:2018mmi}. As discussed in \cref{app:imp_ATRG}, the impurity tenor method can be applied for the second order derivative as well with our initial tensor construction. This can potentially also increase the accuracy.
Our result $\beta_c = 0.6560(3)$ is consistent to the TRG result $\beta_c = 0.656097(1)$ in~\cite{Kuramashi:2018mmi} and the Monte-Carlo result $\beta_c = 0.65608(5)$ in~\cite{Z2monte}.

The calculations show that our approach can successfully be applied to a wide range of systems including gauge theories, and can become a first candidate to investigate a system by means of TRG methods.
The method can be applied to any spin-statistical system which has a finite number of spin degrees of freedom and periodic boundary conditions. We demonstrated this in this section in the case of the $\mathbb{Z}_2$ gauge theory and discuss the generalization and scaling in \cref{app:general}.
Since we do not need a model-specific expansion of the original partition function or integrate out the original variables in our construction, this method can straightforwardly be used for a large class of systems, including gauge theories, to find the tensor network representation of physical quantities.
\section{General form of initial tensors}
\label{app:general}
In this section, we consider the initial tensor construction method with delta functions for general models, including long-range and non-neighboring interactions. We derive the scaling $\phys^{2[\nint + \lstein - 1]}$ for the number of elements of the initial tensors. Here, $\phys{}$ is the dimension of the local Hilbert space, $\nint$ is the number of lattice points of the original, not locally connected tensors representing the partition function. The number of Steiner points $\lstein$ corresponds to the number of lattice points needed to connect isolated regions, as explained later in this section.

\paragraph{Connected long range chain in 1d}

As an example for longer range interactions, we consider a system where each lattice site is coupled to all sites up to a distance of $k$ sites. The partition function can be written as
\be{
Z
=
\sum_{\sigma = 1} ^\phys{}
\prod_{i = 1} ^\N
K_{\sigma_i,\sigma_{i+1},...,\sigma_{i+k}}.\label{eq:appPartitionFunction}
}
The local physical dimension is $\phys$, and $\nint = k+1$ is the number of indices of these initial tensors. See \cref{app:J1J2} for an example of this type.

We apply the decomposition with a delta matrix,
\be{
K_{\sigma_i,\sigma_{i+1},...,\sigma_{i+k}}
=
\sum_{a^{(1)} _{i+1} = 1} ^\phys{}
K_{\sigma_i,\sigma_{i+1},...,a^{(1)} _{i+1}}
\delta_{\sigma_{i+k},a^{(1)} _{i+1}}.
}
Making use of the periodic boundary conditions, we define the new tensor
\begin{align}
&K^{(1)} _{\sigma_i,\sigma_{i+1},...,\sigma_{i+k-1},a^{(1)} _i,a^{(1)} _{i+1}} \nonumber \\
&\equiv
K_{\sigma_i,\sigma_{i+1},...,\sigma_{i+k-1},a^{(1)} _{i+1}}
\delta_{\sigma_{i+k-1},a^{(1)} _{i}},
\end{align}
which leads to the same partition function as the original one if one takes the product of all tensors at different lattice sites and sums over all indices, similar to \cref{eq:appPartitionFunction} but including the new indices $a^{(1)}$.

We can repeat this procedure $k-1$ times to get the local representation
\begin{align}
    &K^{(k-1)} _{\sigma_i,\sigma_{i+1},a^{(1)} _i,a^{(1)} _{i+1},...,a^{(k-1)} _i,a^{(k-1)} _{i+1}} \nonumber \\
    &\equiv K^{(k-1)} _{[\sigma_i a^{(1)} _i \dots a^{(k-1)} _i],[\sigma_{i+1},a^{(1)} _{i+1},\dots,a^{(k-1)} _{i+1}]}. \label{eq:initialTensorsLongRangeSpinChain}
\end{align}
The index dimension of the combined indices $(\sigma\otimes a^{(1)}\otimes...\otimes a^{(k-1)})$ between neighboring points is then $\phys{}^k$, and the initial tensor $K^{(k-1)}$ has $\phys{}^{2k} = \phys{}^{2(\nint-1)}$ elements.

\begin{figure}[htbp]
\begin{center}
 \includegraphics[width=6cm, angle=0]{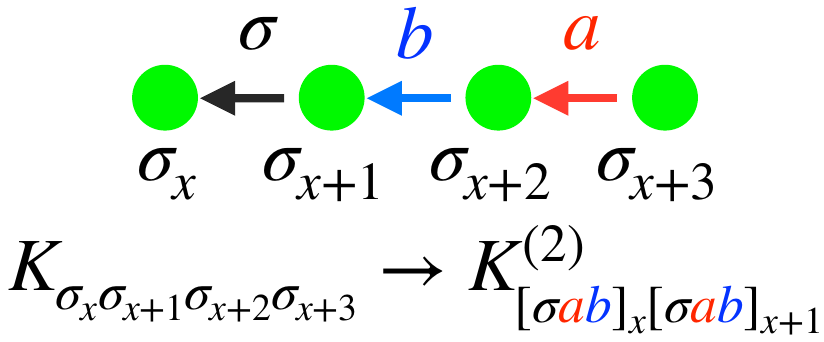}
 \caption{
Schematic picture for the construction of the initial tensor $K^{(2)} _{[\sigma a b]_x[\sigma a b]_{x+1}}$ from the original tensor $K_{\sigma_x\sigma_{x+1}\sigma_{x+2}\sigma_{x+3}}$ for a one-dimensional system with long range interaction.
Green dots represent the indices of the original tensor $K$. Each colored arrow stands for a decomposition with a delta function, which creates new indices in $K^{(2)}$. See main text for details.}
\label{fig:path_kint_1dim}
\end{center}
\end{figure}

\Cref{fig:path_kint_1dim} shows a schematic picture of our method for $k=3$. The original tensor $K_{\sigma_x\sigma_{x+1}\sigma_{x+2}\sigma_{x+3}}$ has four spin variables as indices. These are represented by green dots, and their number is $\nint=4$.
Each decomposition by a delta function creates two new indices and removes the dependence on one spin variable. We represent each such step by a colored arrow. Explicitly, the red arrow removes the dependence on $\sigma_{x+3}$ and creates new indices $a_x$ and $a_{x+1}$. The blue arrow similarly removes the dependence on $\sigma_{x+2}$ and creates new indices $b_x$ and $b_{x+1}$. The black arrow connects nearest neighbors in the original spin indices, and does not correspond to a decomposition.

\paragraph{Disconnected long range interaction in 1d}
The bond size of the tensor network representation in 1d depends only on the maximum interaction distance. For example, we consider a system where the interactions only connect sites at a distance $k$ from each other. The partition function is
\be{
Z
=
\sum_{\sigma = 1} ^\phys{}
\prod_{x=1} ^\N
K_{\sigma_i,\sigma_{i+k}}.
}
Our procedure to construct the initial tensors is similar to the previous example, and leads to the same form of the initial tensors as in \cref{eq:initialTensorsLongRangeSpinChain}. The index size is thus the same, the elements of the tensors differ though. \Cref{app:J1J3} discusses an example of this type of long range interaction.

\begin{figure}[htbp]
\begin{center}
 \includegraphics[width=5cm, angle=0]{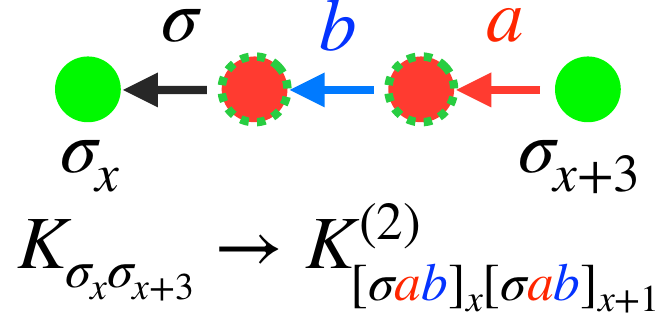}
 \caption{
Schematic picture for the construction of the initial tensor $K^{(2)} _{[\sigma a b]_x[\sigma a b]_{x+1}}$ from the original tensor $K_{\sigma_x\sigma_{x+3}}$ for a one-dimensional system with interaction only between sites with a distance of 3 lattice sites.
Green dots represent the indices of the original tensor $K$. Each colored arrow stands for a decomposition with a delta function, which creates new indices in $K^{(2)}$. Red dots with green outlines are indices which did not exist in the original tensor $K$, but have to be included in our procedure to connect disconnected regions (Steiner points~\cite{RectilinearSteinerTreeProblem}).
}
\label{fig:path_kint_1dim2}
\end{center}
\end{figure}

\Cref{fig:path_kint_1dim2} shows a schematic picture of the procedure. The red arrow removes the dependence on $\sigma_{x+k}$ but introduces a dependence on the site $x+k-1$, which is denoted as a red dot in our graphical notation. This new dependence is removed by the blue arrow. The green outlines indicate that the original tensor did not depend on these sites. The number of arrows is the same as in the previous example, and thus the resulting tensor has the same dimensions.

We define the number of arrows, which corresponds to the number of decompositions in our method, as $\ndec$. The initial tensor can then be represented as a $(\phys{}^{\ndec} \times \phys{}^{\ndec})$ matrix $K^{(k-1)}$.
This can also be expressed in terms of the number of original spin values (green dots) $\nint$ and the number of generated Steiner points (red dots) $\lstein$. The latter are needed to connect disconnected regions of the lattice, and are the points with green outlines in \cref{fig:path_kint_1dim2}. In the case discussed here, the new tensor has size $(\phys{}^{\nint - 1 + \lstein} \times \phys{}^{\nint - 1 + \lstein})$, and thus has $\phys{}^{2(\nint - 1 + \lstein)}$ elements.

\paragraph{Higher dimensions}
\begin{figure}[htbp]
\begin{center}
 \includegraphics[width=8cm, angle=0]{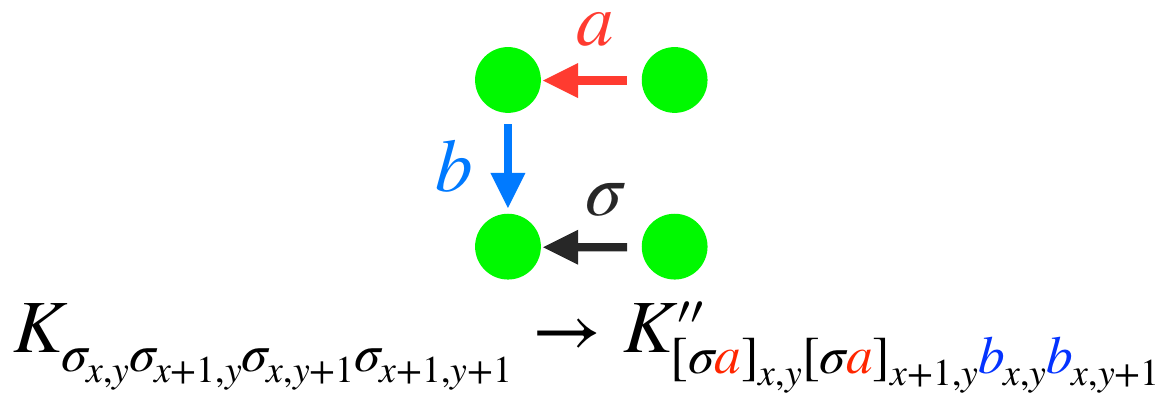}
 \caption{
Schematic picture of the decompositions and initial tensor indices for the  $J_1-J_2$ Ising model (see \cref{app:J1J2}) or a plaquette term in two dimensions. See \cref{fig:path_kint_1dim} for symbology.
}
\label{fig:path_J1J2}
\end{center}
\end{figure}

\begin{figure}[htbp]
\begin{center}
 \includegraphics[width=8cm, angle=0]{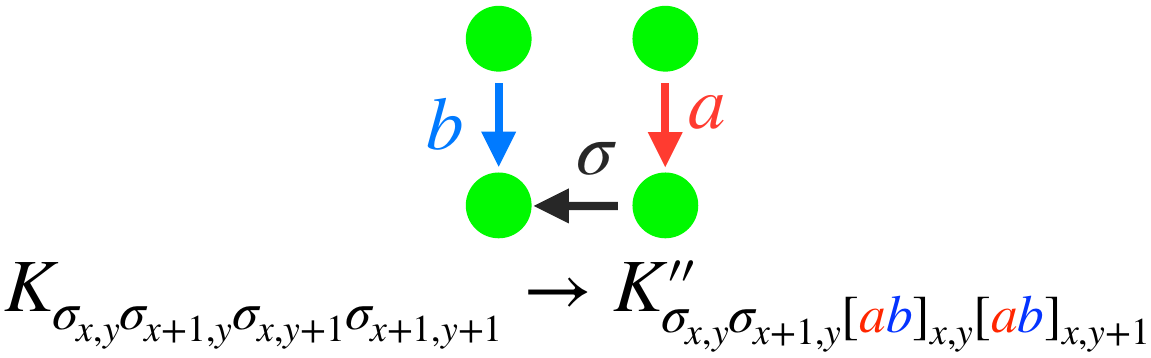}
 \caption{
Alternative to \cref{fig:path_J1J2} for constructing the initial tensor. See \cref{fig:path_kint_1dim} for symbology. The number of arrows and therefore the number of tensor elements remains the same, but the shape of the initial tensor differs.
}
\label{fig:path_J1J2_alt}
\end{center}
\end{figure}

The same scaling $\phys{}^{2(\nint - 1 + \lstein)}$ holds in higher dimensions as well. However, $\nint$ typically grows in higher dimensions because interactions happen in more directions. We can use the graphical notation again, as shown for example in \cref{fig:path_J1J2}. Arrows are introduced such that a path arises from all sites that take part in the interaction to the origin. Each arrow in a given spatial direction in the lattice contributes a factor $d$ in the bond size of the index for this direction in the constructed tensor.
Note, however, that the choice of arrows is not unique anymore in more than one dimension. For example, \cref{fig:path_J1J2_alt} shows an alternative way to connect the tensors compared to \cref{fig:path_J1J2}. The constructed tensor has the same number of elements in this case, but the dimensions of the individual indices differ.

\begin{figure}[htbp]
\begin{center}
 \includegraphics[width=8cm, angle=0]{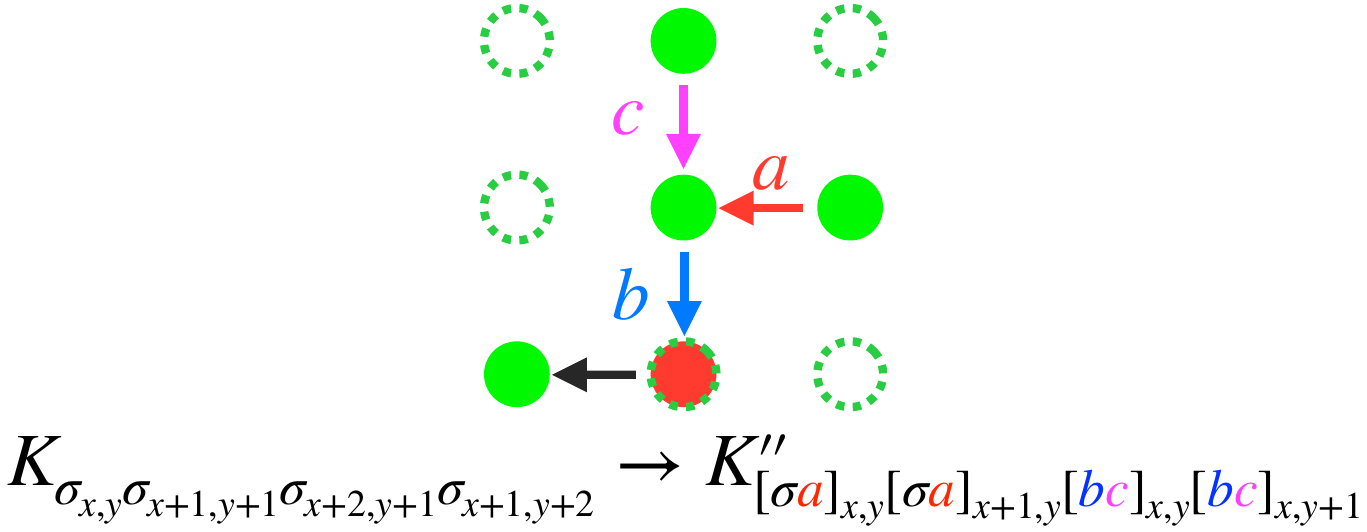}
 \caption{
Schematic picture of the decompositions for constructing the initial tensors for a more complicated interaction structure. See \cref{fig:path_kint_1dim2} for symbology. Green outlined dots are indices that are not present in the original tensor. However, the red dot with a green outline needs to be included to connect disconnected regions. The model is explained in the main text.
}
\label{fig:path_RSTM}
\end{center}
\end{figure}

Finally, we discuss the example in \cref{fig:path_RSTM} where isolated regions arise. The nearest neighbors of the lower left site do not take part in the interaction, which is symbolized by dashed red outlines of these sites. To form a connected graph, at least one isolated point has to be included. Finding the minimum number of arrows in our graphical representation is a well known problem in graph theory, known as the rectilinear Steiner tree problem~\cite{RectilinearSteinerTreeProblem}. The graph in \cref{fig:path_RSTM} has $n_x = 2$ arrows in the x-direction, $n_y = 2$ arrows in the y-direction and one isolated point. Thus, the constructed tensor has the dimensions $(\phys{}^{n_x} \times \phys{}^{n_x} \times \phys{}^{n_y} \times \phys{}^{n_y})$ and $\phys{}^{2[\nint + \lstein - 1]} = \phys{}^{2[4 + 1 - 1]} = \phys{}^8$ elements.

The connectivity of \cref{fig:path_RSTM} allows for various types of interactions. It can express nearest neighbor interactions in positive and negative x- and y-directions, next-to-nearest neighbor interactions (diagonal), and next-to-nearest neighbor interactions (one site up, two sites in y-direction, or two sites up and one site in y-direction). Moreover, three- and four-site interactions are possible. The most generic form of a spin model of this type has 12 parameters. Even such an involved model can be expressed with an initial tensor of moderate dimensions $(4 \times 4 \times 4 \times 4)$ for $\phys=2$.
The explicit form of possible interactions for the graph in \cref{fig:path_RSTM} is:
\beqnn{
&&
K_{\sigma_{x,y}\sigma_{x+1,y+1}\sigma_{x+1,y+2}\sigma_{x+2,y+1}}\nonumber\\
&=&
e^{h(\sigma_{x,y}+\sigma_{x+1,y+1}+\sigma_{x+1,y+2}+\sigma_{x+2,y+1})}\nonumber\\
&\times&
e^{J_1 ^{(x)}\sigma_{x+1,y+1}\sigma_{x+2,y+1}+J_1 ^{(y)}\sigma_{x+1,y+1}\sigma_{x+1,y+2}}\nonumber\\
&\times&
e^{J_2 ^{(1)}\sigma_{x,y}\sigma_{x+1,y+1}+J_2 ^{(2)}\sigma_{x+1,y+2}\sigma_{x+2,y+1}}\nonumber\\
&\times&
e^{g_3 ^{(1)}\sigma_{x,y}\sigma_{x+2,y+1}+g_3 ^{(2)}\sigma_{x,y}\sigma_{x+1,y+2}}\nonumber\\
&\times&
e^{t_8\sigma_{x,y}\sigma_{x+2,y+1}\sigma_{x+1,y+2}}\nonumber\\
&\times&
e^{t_6 ^{(1)}\sigma_{x,y}\sigma_{x+1,y+1}\sigma_{x+2,y+1}
+t_6 ^{(2)}\sigma_{x,y}\sigma_{x+1,y+1}\sigma_{x+1,y+2}}\nonumber\\
&\times&
e^{t_4 \sigma_{x+1,y+1}\sigma_{x+2,y+1}\sigma_{x+1,y+2}
}\nonumber\\
&\times&
e^{f\sigma_{x,y}\sigma_{x+1,y+1}\sigma_{x+2,y+1}\sigma_{x+1,y+2}
}.
}
\paragraph{Multi-flavour systems}
So far we only considered one-flavour systems, but the ideas can be generalized easily to multi-flavour systems. For example, the degree of freedom of the $\mathbb{Z}_2$ model can be located on the links pointing from $\hat{r}$ to $\hat{r} + \hat{\mu}$, where $\hat{r}$ is a coordinate and $\hat{\mu}\in \{\hat{x},\hat{y},\hat{z} \}$ is a unit vector in one of the three directions. This is indicated in \cref{fig:plaquette}. We can also localize each such gauge degrees of freedom at the node with position $\hat{r}$. Then, at each node, an additional degree of freedom arises for the three cases $\hat{\mu} = \hat{x}$, $\hat{\mu} = \hat{y}$, $\hat{\mu} = \hat{z}$. The connectivity is then the same as for the $J_1-J_2$ model, but with three distinct flavours. There is no Steiner point, $\lstein = 0$, and the degrees of freedom in each direction is two, such that the number of elements of the initial tensor is $\left(2^3\right)^{2(3 - 1)} = 2^{12} = (4^3)^2$. The initial tensors can be formed by a $(4\times 4\times 4\times 4\times 4\times 4)$ tensor as shown in the main text in \cref{sec:Z2}.

%
%
%
%
%
%
%
%
\section{Conclusion\label{sec:Conclusion}}
In this paper we introduced a simple construction of a tensor network representing a partition function. 
By inserting a delta function and redefining the tensors, we can construct a locally connected tensor network for any theory with periodic boundary conditions in any dimension. This network can then be coarse-grained with TRG methods to calculate the partition function and observables for translationally invariant systems.

In a general case, a partition function can be represented by an initial tensor with $\phys^{2(\nint - 1  + \lstein)}$ elements (see \cref{app:general}). Here, $\phys{}$ is the dimension of the local degrees of freedom, and $\nint$ is the number of indices of the original tensor, which did not form a locally connected tensor network. If disconnected regions exist in the interactions, $\lstein$ corresponds to the Steiner points~\cite{RectilinearSteinerTreeProblem} that are needed to connect these regions.

We demonstrated the applicability of our method in a one-dimensional spin system with multiple interaction terms as a simple example.
We extended the method to two-dimensions and investigated the initial tensor dependence of the TRG method.
The accuracy of these methods highly depends on the initial tensors and on the details of the TRG method. A high sensitivity was found in the original HOTRG.
Our results suggest that one should use symmetric initial tensors for this method.
We conclude that the initial tensor influences the numerical accuracy significantly depending on the TRG method, and should be chosen carefully for reliable calculations.
We found that symmetric initial tensors lead to better results for many coarse-graining methods, and we calculated a symmetric representation for the two-dimensional Ising model based on our initial tensor construction.

Moreover, we showed that the initial tensor dependence can be eliminated by applying the ideas of the boundary TRG method to HOTRG.
In general, any TRG method, such as ATRG and MDTRG, that makes use of isometries to form the new indices of the coarse grained tensors, has a strong initial tensor dependence. We showed, however, how these methods can be modified slightly to use squeezers instead of isometries, as introduced in the boundary TRG~\cite{boundaryHOTRG}. This way, the dependence on initial tensors and their symmetries can be removed, which makes the algorithms more resilient against systematic errors coming from an interplay between the choice of initial tensors and the coarse graining method.

The precision of TRG algorithms also depends on the type of index-exchange between coarse-graining steps. There are several possibilities to alternate between coarse-graining in $x-$ and $y-$ directions. We showed that systematic errors can accumulate with the wrong type of index exchange and discussed the optimal choice for different coarse-graining methods.

We further applied our tensor construction to the $\mathbb{Z}_2$ gauge theory in three-dimensions without gauge-fixing.
We neither need to consider any expansion nor do we have to integrate out original variables.
The results for the free energy and the specific heat with our simple tensor construction were consistent with TRG calculations using expansions and gauge-fixing, and with Monte-Carlo simulations.
For the $\mathbb{Z}_2$ gauge theory, our construction resulted in an accuracy of the free energy comparable to that of the usual construction by an expansion. We used the impurity tensor method for the calculation of derivatives. Our initial tensor construction can be applied with site-dependent tensors as well and is thus particularly suitable for the impurity method.

Summarizing, the initial tensor construction presented in this work is a way to translate the partition function to a locally connected tensor network. The approach is simple and can be applied to various systems, without relying on model-specific expansions. Moreover, we found a worrisome dependence of HOTRG-like methods (isometric ATRG, MDTRG, HOTRG) on the choice of initial tensors. Even if different choices are mathematically equivalent, the truncation procedures of the coarse graining steps introduce systematic errors. The previously mentioned methods should therefore only be used in their original form for symmetric initial tensors. However, we found that the methods can be made resilient against errors from the choice of initial tensors by using the ideas of the boundary TRG. With this, or by choosing alternative coarse graining algorithms like ATRG, our initial tensor construction leads to a similar accuracy as other construction methods, making it a simple and powerful tool for TRG calculations.

\section*{Acknowledgments}
We would like to thank Shinji Takeda and Daisuke Kadoh for discussions.
This work was supported by JSPS KAKENHI Grant Number 24K17059.

\appendix

%
%
%
%
%
%
%
%
%
\section{J1-J2 Ising model}
\label{app:J1J2}
In this appendix, we discuss how our method can be used to construct the tensor network representation of the $J_1-J_2$ Ising model.
The system with $N$ sites can be described by the partition function 
\begin{align}
Z
=&
\sum_{\sigma = \pm 1}
\prod_{x,y = 1} ^N
e^{\frac{J_1}{2}(
\sigma_{x,y}\sigma_{x+1,y}
+ \sigma_{x,y}\sigma_{x,y+1})}\nonumber\\
&\times
e^{\frac{J_1}{2}(\sigma_{x,y+1}\sigma_{x+1,y+1}
+ \sigma_{x+1,y}\sigma_{x+1,y+1})}\nonumber\\
&\times
e^{J_2(\sigma_{x,y}\sigma_{x+1,y+1} + \sigma_{x,y+1}\sigma_{x+1,y}) } \\
=&
\sum_{\sigma = \pm 1}
\prod_{x,y = 1} ^N
K^{(J_1J_2)} _{\sigma_{x,y},\sigma_{x+1,y},\sigma_{x,y+1},
\sigma_{x+1,y+1}},
\end{align}
with the spin indices $\sigma_{x,y}$ at sites $\{x,y\}$ and coupling constants $J_1$ and $J_2$.
By setting the coupling $J_1 < 0$ and $J_2 > 0$, frustrated systems can be studied in this model.

The representation through the tensor $K^{(J_1J_2)}$ is not a two-dimensional locally connected tensor network.
We can construct such a network by inserting delta functions.
First, we split the next-nearest neighbor spin $\sigma_{x+1,y+1}$ from the tensor:
\begin{align}
&
K^{(J_1J_2)} _{\sigma_{x,y},\sigma_{x+1,y},\sigma_{x,y+1},
\sigma_{x+1,y+1}}\nonumber\\
&=
\sum_{a = \pm 1}
K^{(J_1J_2)} _{\sigma_{x,y},\sigma_{x+1,y},\sigma_{x,y+1},
a_{x+1,y}}
\delta^{a_{x+1,y}}   _{\sigma_{x+1,y+1}}.
\end{align}
With this, we define a new tensor $K^{'(J_1J_2)}$,
\begin{align}
&
K^{'(J_1J_2)}  _{\sigma_{x,y},\sigma_{x+1,y},\sigma_{x,y+1},a_{x,y},a_{x+1,y}}\nonumber\\
&\equiv
K^{(J_1J_2)} _{\sigma_{x,y},\sigma_{x+1,y},\sigma_{x,y+1},a_{x+1,y}}
\delta^{a_{x,y}}   _{\sigma_{x,y+1}}.
\end{align}

As a next step, we split the index $\sigma_{x,y+1}$ from the tensor:
\begin{align}
&
K^{'(J_1J_2)} _{\sigma_{x,y},\sigma_{x+1,y},\sigma_{x,y+1},a_{x,y},a_{x+1,y}}\nonumber\\
&=
\sum_{b = \pm 1}
K^{'(J_1J_2)} _{\sigma_{x,y},\sigma_{x+1,y},b_{x,y+1},a_{x,y},a_{x+1,y}}
\delta^{b_{x,y+1}} _{\sigma_{x,y+1}},
\end{align}
and define the new tensor $K^{''(J_1J_2)}$:
\begin{align}
&
K^{''(J_1J_2)} _{\sigma_{x,y},\sigma_{x+1,y},a_{x,y},a_{x+1,y},b_{x,y},b_{x,y+1}}\nonumber\\
&\equiv
K^{'(J_1J_2)} _{\sigma_{x,y},\sigma_{x+1,y},b_{x,y+1},a_{x,y},a_{x+1,y}}
\delta^{b_{x,y}} _{\sigma_{x,y}}
\\
&=
K^{(J_1J_2)} _{\sigma_{x,y},\sigma_{x+1,y},b_{x,y+1},a_{x+1,y}}
\delta^{a_{x,y}} _{b_{x,y+1}}
\delta^{b_{x,y}} _{\sigma_{x,y}}
\\
&=
e^{\frac{J_1}{2}(
\sigma_{x,y}\sigma_{x+1,y}
+ \sigma_{x,y}b_{x,y+1})}\nonumber\\
&\times
e^{\frac{J_1}{2}(b_{x,y+1}a_{x+1,y}
+ \sigma_{x+1,y}a_{x+1,y})}\nonumber\\
&\times
e^{J_2(\sigma_{x,y}a_{x+1,y} + b_{x,y+1}\sigma_{x+1,y}) }\nonumber\\
&\times
\delta^{a_{x,y}} _{b_{x,y+1}}
\delta^{b_{x,y}} _{\sigma_{x,y}}.
\label{eq:J1J2}
\end{align}

We combine the $\sigma$ and $a$ indices to form new bonds in $x$-direction: at position $x,y$, the new index is $[\sigma a]_{x,y} \equiv \sigma_{x,y}\otimes a_{x,y} = (\sigma_{x,y},a_{x,y})$. Finally, the local tensor representation in terms of $K''$ is
\begin{equation}
Z
=
\sum_{[\sigma a]}
\sum_{b=\pm 1}
\prod_{x,y = 1} ^N
K^{''(J_1J_2)} _{[\sigma a]_{x,y},[\sigma a]_{x+1,y},b_{x,y},b_{x,y+1}}.
\label{eq:J1J2_last}
\end{equation}
The indices of this representation are independent of each other, and this initial tensor can be used for TRG coarse-graining.

We note that other representations can also be starting points for our procedure, as long as the contraction of the initial tensor network corresponds to the same partition function. For example, we can substitute $K$ in \cref{eq:J1J2} by $K^{(0)}$,
\begin{align}
&
K^{(0)} _{\sigma_{x,y},\sigma_{x+1,y},\sigma_{x,y+1},
\sigma_{x+1,y+1}}
\nonumber\\
&=
e^{{J_1}(
\sigma_{x,y}\sigma_{x+1,y}
+ \sigma_{x,y}\sigma_{x,y+1})}\nonumber\\
&\times
e^{J_2(\sigma_{x,y}\sigma_{x+1,y+1} + \sigma_{x,y+1}\sigma_{x+1,y}) }.
\end{align}
In any case, the tensor construction reproduces the original partition function if all indices of the network are contracted.

Our procedure results in an alternative representation of the partition function to those studied in~\cite{J1J2delta,J1J2HOTRG}. The authors of~\cite{J1J2HOTRG} state that physical quantities depend strongly on the choice of initial tensors and, for finite lattices, on the boundary conditions implemented by the tensor network representation. Additional candidates for initial tensors can therefore be helpful to find the most accurate representation for a given algorithm and system size.
%
%
%
%
%
%
%
%
%
%

%
%
%
%
%
%
%
%
%
\section{J1-J3 Ising model}
\label{app:J1J3}
As a kind of third-nearest neighbor Ising model, we discuss the $J_1-J_3$ Ising model, which is also called biaxial next-nearest neighbor Ising model.
The partition function is
\begin{align}
Z
=&
\sum_{\sigma = \pm1}
\prod_{x,y = 1} ^N
e^{J_1(
\sigma_{x,y}\sigma_{x+1,y}
+ \sigma_{x,y}\sigma_{x,y+1})}\nonumber\\
&\times
e^{J_3(\sigma_{x,y}\sigma_{x+2,y} + \sigma_{x,y}\sigma_{x,y+2}) } \\
=&
\sum_{\sigma = \pm 1}
\prod_{x,y = 1} ^N
K^{(J_1J_3)} _{\sigma_{x,y},\sigma_{x+1,y},\sigma_{x+2,y},\sigma_{x,y+1},
\sigma_{x,y+2}}
\end{align}
We split the next-next-nearest spins $\sigma_{x+2,y}$ and $\sigma_{x,y+2}$ from the tensor,
\begin{align}
&
K^{(J_1J_3)} _{\sigma_{x,y},\sigma_{x+1,y},\sigma_{x+2,y},\sigma_{x,y+1},
\sigma_{x,y+2}}\nonumber\\
&=
\sum_{a,b = \pm 1}
K^{(J_1J_3)} _{\sigma_{x,y},\sigma_{x+1,y},a_{x+1,y},\sigma_{x,y+1},
b_{x,y+1}}\nonumber\\
&\times
\delta_{a_{x+1,y},\sigma_{x+2,y}}
\delta_{b_{x,y+1},\sigma_{x,y+2}},
\end{align}
and define the tensor $K^{'(J_1J_3)}$ with shifted delta functions:
\begin{align}
&
K^{'(J_1J_3)} _{\sigma_{x,y},\sigma_{x+1,y},\sigma_{x,y+1},a_{x,y},a_{x+1,y},
b_{x,y},b_{x,y+1}}\nonumber\\
&=
K^{(J_1J_3)} _{\sigma_{x,y},\sigma_{x+1,y},a_{x+1,y},\sigma_{x,y+1},
b_{x,y+1}}\nonumber\\
&\times
\delta_{a_{x,y},\sigma_{x+1,y}}
\delta_{b_{x,y},\sigma_{x,y+1}}.
\end{align}
Similarly, we split $\sigma_{x,y+1}$,
\begin{align}
&
K^{'(J_1J_3)} _{\sigma_{x,y},\sigma_{x+1,y},\sigma_{x,y+1},a_{x,y},a_{x+1,y},
b_{x,y},b_{x,y+1}}\nonumber\\
&=
\sum_{c = \pm 1}
K^{'(J_1J_3)} _{\sigma_{x,y},\sigma_{x+1,y},c_{x,y+1},a_{x,y},a_{x+1,y},
b_{x,y},b_{x,y+1}}\nonumber\\
&\times
\delta_{c_{x,y+1},\sigma_{x,y+1}},
\end{align}
and define the tensor $K^{''(J_1J_3)}$:
\begin{align}
&
K^{''(J_1J_3)} _{\sigma_{x,y},\sigma_{x+1,y},c_{x,y},c_{x,y+1},a_{x,y},a_{x+1,y},
b_{x,y},b_{x,y+1}}\nonumber\\
&=
K^{'(J_1J_3)} _{\sigma_{x,y},\sigma_{x+1,y},c_{x,y+1},a_{x,y},a_{x+1,y},
b_{x,y},b_{x,y+1}}\nonumber\\
&\times
\delta_{c_{x,y},\sigma_{x,y}}.
\end{align}

We define the new indices in $x$-direction at position $x,y$ as $[\sigma a]_{x,y} = (\sigma_{x,y},a_{x,y})$, and $[cb]_{x,y} = (c_{x,y},b_{x,y})$ for the $y$-direction. We finally obtain the locally connected tensor network representation
\begin{equation}
Z
=
\sum_{[\sigma a],[c,b]}
\prod_{x,y = 1} ^N
K^{''(J_1J_3)} _{[\sigma a]_{x,y}[\sigma a]_{x+1,y} [cb]_{x,y}[cb]_{x,y+1}}.
\end{equation}

Compared to the nearest-neighbor Ising model (see \cref{eq:2d_ising_Kdelta}) and the $J_1-J_2$ Ising model (see \cref{eq:J1J2_last}), the $J_1-J_3$ Ising model is represented by an initial tensor with a larger bond dimension of the combined indices. This is a typical property for models with longer range interactions: they require a larger number of the decompositions, and thus create additional new indices in the locally connected tensor network representation. When these indices are combined, the new bonds have a larger bond dimension. See \cref{app:general} for a general discussion of the scaling behavior.

%
%
%
%
%
%
%
%
%
\section{Symmetry of the initial tensor}
\label{app:sym_TRG}
In order to investigate the initial tensor dependence of various TRG algorithms for the two-dimensional Ising model in \cref{sec:IsingModel}, we consider the symmetrized tensor $K^{\mathrm{(sym)}}$ as a variant of $K^{(\mathrm{delta})}$.
The partition function in \cref{eq:TN2dim} does not change if we redefine the initial tensor as
\be{
	K^\mathrm{(sym)} _{XX'YY'}
		\equiv
        \sum_{kk'll'}
		A_{Xk}
		A_{X'k'}
		K^{(\mathrm{delta})} _{kk'll'}
		A_{lY} ^{-1}
		A_{l'Y'} ^{-1}.
}
This reconstructed tensor can be made symmetric under swapping of the two indices $K_{abcd} = K_{\{abcd\}}$ if we choose $A$ in the right way.

\begin{figure}[htbp]
\begin{center}
 \includegraphics[width=8cm, angle=0]{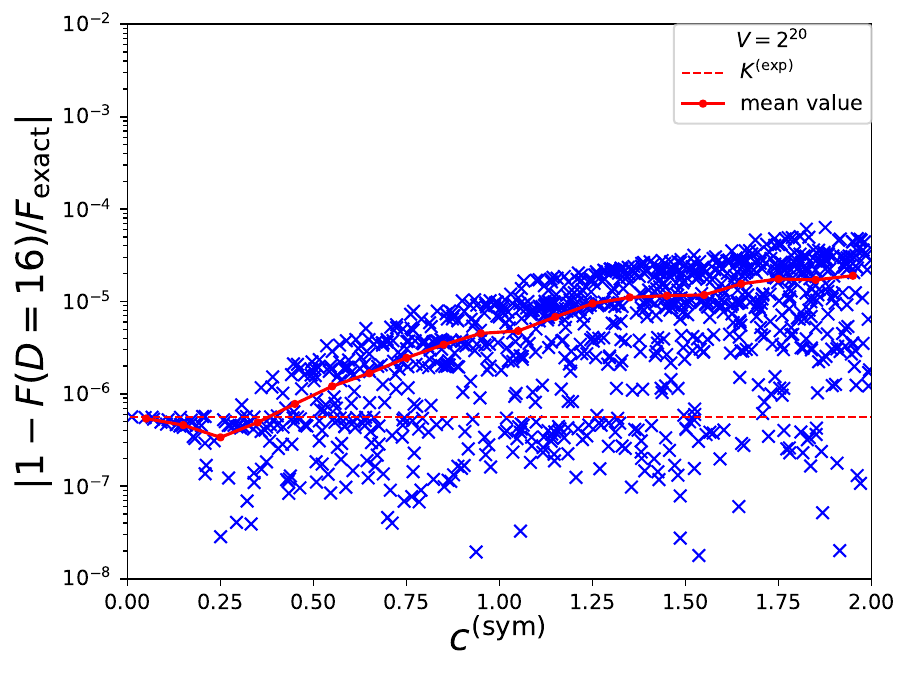}
 \includegraphics[width=8cm, angle=0]{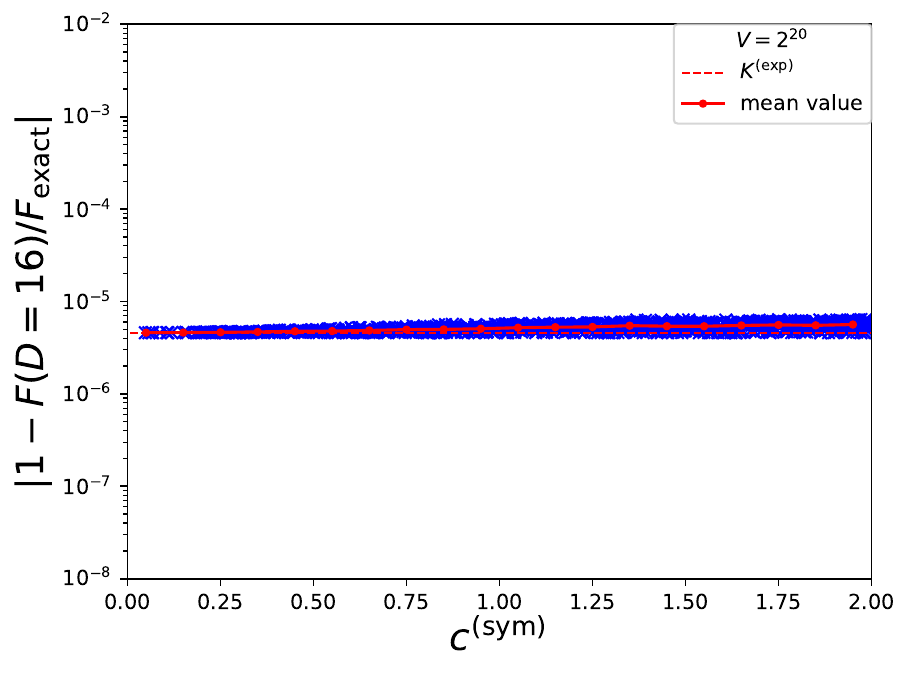}
 \caption{
 Dependence of TRG methods on the symmetry $c^\mathrm{(sym)}$ of the initial tensors (see \cref{eq:symm_costfunction}) for the two-dimensional critical Isinig model. Shown is the relative error of the free energy for 1000 different initial tensors, which are obtained from an incomplete symmetrization process of the asymmetric tensor $K^{(\mathrm{delta})}$. See main text for details. The dotted lines are results with the symmetric tensor $K^{(\mathrm{exp})}$ with $c^\mathrm{(sym)}=0$. Dots are mean values in a bin of width 0.1. (Upper panel): HOTRG with strong symmetry dependence; (Lower panel:) TRG with no significant dependence on the initial tensors.
}
\label{fig:HOTRG_rand}
\end{center}
\end{figure}

Several methods are possible to find a suitable $A$ to make $K^\mathrm{(sym)} _{XX'YY'}$ a symmetric tensor. We apply a numerical optimization starting from a random matrix.
This matrix is optimized element-wise to minimize the cost function 
\be{
	c^\mathrm{(sym)}
	\equiv
	\sum_{xx'yy'}
	|
	K_{XX'YY'}
	-
	K_{\{xx'yy'\}}
	|.
 \label{eq:symm_costfunction}
}
In each optimization step we change a matrix element by the step size $\Delta \sim 10^{0\sim -9}$ and consider $A_{kl}' = A_{kl} \pm \Delta$. We choose either $A'$ or $A$ for the next step and accept or reject the change, depending on which of the two has the lower cost function. We sweep several times through all matrix elements and decrease $\Delta$ if all $A'$ get rejected.
Because the optimization can be stuck in local minima, we repeat the optimization with different randomly initialized matrices $A$, until we find 1000 matrices with $c^\mathrm{(sym)} \leq 2$. The partition function is calculated with HOTRG and the results are shown in \cref{fig:HOTRG_rand} for all outcomes of this optimization. 
We found a tensor $K^\mathrm{(sym)} _{XX'YY'}$ with a cost function $c^\mathrm{(sym)}$ smaller than $10^{-1}$. The explicit form of $K^\mathrm{(sym)}$ for $h = 0$, $g = 1$ is given in \cref{eq:Ksym}.
Note that $K^\mathrm{(sym)} \neq K^\mathrm{(\mathrm{exp})}$, although $K^\mathrm{(\mathrm{exp})}$ is also a symmetric tensor.

%
%
%
%

The free energy calculated with the symmetrized tensor $K^\mathrm{(sym)}$ is shown in \cref{fig:TRG_HOTRG} for HOTRG, and the accuracy is similar to a calculation with $K^\mathrm{(\mathrm{exp})}$. This shows that a symmetrization of the initial tensor $K^{(\mathrm{delta})}$ can improve the results for symmetry-dependent TRG methods like HOTRG.

Furthermore, we study the $c^\mathrm{(sym)}$ dependence of the TRG and HOTRG methods in \cref{fig:HOTRG_rand}. The results clearly show that the HOTRG method becomes less precise and accurate when the initial tensors are less symmetric and $c^\mathrm{(sym)}$ is large. In contrast to this, TRG shows almost no dependence on the symmetry behavior of the initial tensors.


We list the explicit representation of the symmetrized initial tensor for the two-dimensional Ising model in the following.
For each $\{\sigma, a\}$ we define the combined index $[\sigma a]$.
The indices are ordered as $[- -]=0,[+ -]=1,[- +]=2,[+ +]=3$.
Then, the symmetrized tensor $K^\mathrm{(sym)} _{[\sigma_{x,y} a_{x,y}][\sigma_{x+1,y}a_{x,y+1}]}$ is 
\begin{align}
    K^\mathrm{(sym)} _{00} =& 2.48037458878,\nonumber\\
    K^\mathrm{(sym)} _{01} = K^\mathrm{(sym)} _{02} =& 0.167834510235,\nonumber\\
    K^\mathrm{(sym)} _{10} = K^\mathrm{(sym)} _{20} =& 0.166746023749,\nonumber\\
    K^\mathrm{(sym)} _{11} = K^\mathrm{(sym)} _{12} = K^\mathrm{(sym)} _{21} = K^\mathrm{(sym)} _{22} =& 0.334196191574,\nonumber\\
    K^\mathrm{(sym)} _{13} = K^\mathrm{(sym)} _{23} =& 0.749091024240,\nonumber\\
    K^\mathrm{(sym)} _{31} = K^\mathrm{(sym)} _{32} =& 0.749047098416,\nonumber\\
    K^\mathrm{(sym)} _{03}=& 0.334224186621,\nonumber\\
    K^\mathrm{(sym)} _{30}=& 0.334168680654,\nonumber\\
    K^\mathrm{(sym)} _{33}=& 1.67966015282.\label{eq:Ksym}
\end{align}
Note that this initial tensor is not exactly symmetric: to achieve $c^\mathrm{(sym)} = 0$, the relation $K^\mathrm{(sym)} _{ab} = K^\mathrm{(sym)} _{ba}$ must hold for any index.
However, the symmetry is sufficient for a reliable coarse graining with sufficient accuracy as can be seen in \cref{fig:TRG_HOTRG}.
%
%
%
%
%
%
%
%
%
%

%
%
%
%
%
%
%
%
%
\section{Boundary TRG method}
\label{app:b_TRG}

The boundary TRG method was originally introduced for open boundary systems to take into account the boundary effect in the coarse graining step.
In this appendix, we present a generalization of the original HOTRG method~\cite{HOTRG} using the boundary TRG technique~\cite{boundaryHOTRG}, which removes the dependence on the symmetry properties of the initial tensors. The idea can be generalized to other tensor renormalization methods.

\begin{figure}[htbp]
\begin{center}
    \begin{align}
        C_U =&
            \parbox{0.9\linewidth}{
                \includegraphics[width=\linewidth]{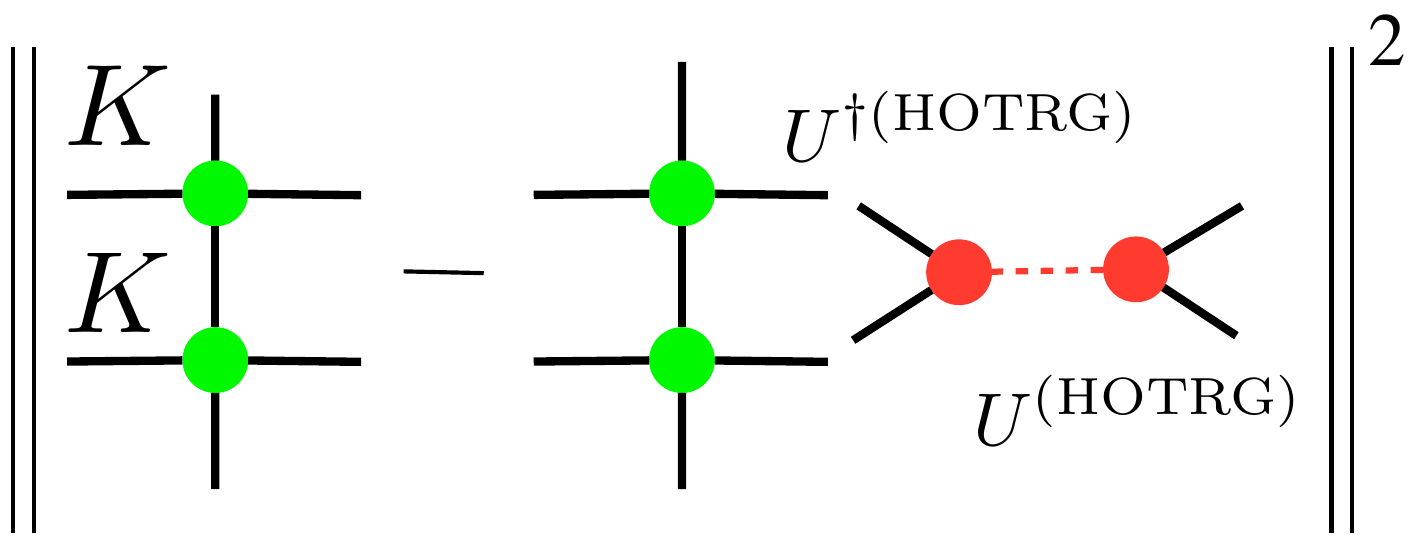}
            }
            \nonumber
            \\
        C_V =&
            \parbox{0.9\linewidth}{
                \includegraphics[width=\linewidth]{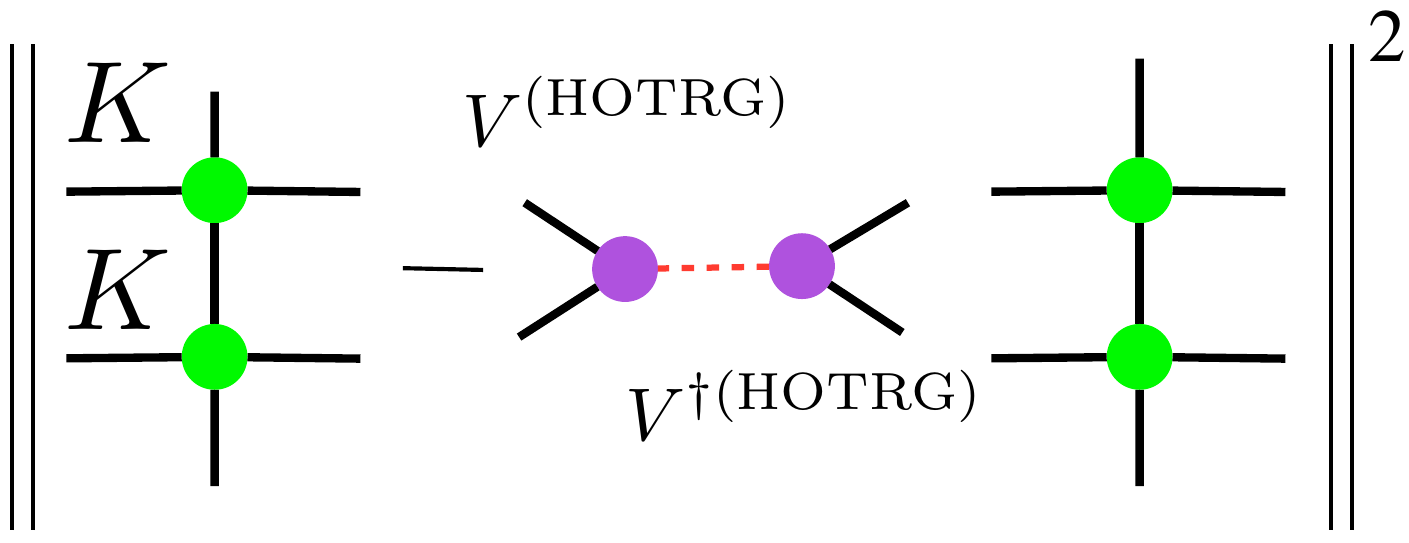}
            }
            \nonumber
    \end{align}
 \caption{
    Cost functions of the isometries $U^\mathrm{(HOTRG)}$ and $V^\mathrm{(HOTRG)}$ for HOTRG.
}
\label{fig:Iso_of_HOTRG}
\end{center}
\end{figure}

The difference between common TRG methods like HOTRG and boundary TRG is the truncation method in the coarse-graining step.
In the original HOTRG, the isometries $U^{\mathrm{(HOTRG)}}$ and $V^{\mathrm{(HOTRG)}}$, which minimize the cost function in \cref{fig:Iso_of_HOTRG}, are both calculated.
The isometries are found by truncated SVDs with singular values $\lambda ^{(U)}$ and $\lambda ^{(V)}$. 
For example, for $\lambda ^{(U)}$:
\begin{align}
\sum_{x_1,x_2,y,y^t,y_1 ',y_2}&
K_{x_1 y^t {x_1 '} ^t y_1 '} ^*
K_{x_2 y_2 {x_2 '} ^t y^t  } ^*
K_{x_1 y   x_1 '    y_1 '}
K_{x_2 y_2 x_2 '    y    }\nonumber\\
\simeq
\sum_{a,b} ^D
&U^{*(\mathrm{HOTRG})} _{a {x_1 ' }^t {x_2 ' }^t}
\left(
\lambda ^{(U)}
\right)^2_{ab}
U^{(\mathrm{HOTRG})} _{b x_1 'x_2 '}.
\label{eq:Isohotrg}
\end{align}

Here, $x_i$ ($x'_i$) are the indices that connect the tensor $K$ to its nearest neighbor to the left (right). Accordingly, $y_i$ ($y'_i$) connects to the next tensor below (above). Upper labels $t$ as in $x^t$ indicate that these bonds connect conjugate tensors.
For brevity, we drop the indices in the following and use a shorthand notation like $K^\dagger K^\dagger KK \simeq U^\mathrm{\dagger (HOTRG)}\left(\lambda^{(U)}\right)^2 U^\mathrm{(HOTRG)}$.\footnote{On the notation for the SVD used here: a Hermitian matrix $M$ can be written as $M = A^\dagger A$. With the SVD $A = U \lambda V$, we can decompose $M$ as $M = V^\dagger \lambda U^\dagger U \lambda V = V^\dagger \lambda^2 V$. In actual calculations, we decompose $M$ in an SVD as $M = U_M \lambda_M V_M$ and identify $V=V_M=U_M^\dagger$, $\lambda^2 = \lambda_M$. We use the names $U$ and $V$ interchangeably for isometries. Typically, we label isometries as $U$, and call them $V$ whenever they have to be distinguished from a given $U$ because they act on different indices of a tensor. Furthermore, we do not put any daggers $\dagger$ on tensors in SVDs. With this convention, isometries are always applied in the form $U^\dagger$ or $V^\dagger$ to the tensors when indices shall be combined and truncated.
\label{fn:SVDhermitean}} The indices can be reconstructed from the corresponding diagrams.

In the cost function $C_U$ in \cref{fig:Iso_of_HOTRG}, $U^{\dagger(\mathrm{HOTRG})}$ is applied to the right indices of the $K$ tensors. Instead, one can also apply an isometry to the left indices. The corresponding cost function $C_V$ is minimized by $V^{\dagger(\mathrm{HOTRG})}$ as an isometry.
In general, the isometries $U^{(\mathrm{HOTRG})}$ and $V^{(\mathrm{HOTRG})}$ are different. In the usual HOTRG algorithm, the cost functions $C_U$ and $C_V$ are computed by summing the squared truncated singular values $\left( \lambda_{>D} ^{(U)} \right)^2$ and $\left( \lambda_{>D} ^{(V)} \right)^2$ in both cases. Then, the isometry which corresponds to the smaller cost function is chosen for the truncation step. This introduces a systematic error, which favors one direction (left or right in \cref{fig:Iso_of_HOTRG}) in the truncation. In the case of symmetric initial tensors, $C_U$ and $C_V$ are the same in each step and thus no choice is needed. Since no direction is favored in this case, the algorithm is more suited for symmetric initial tensors than for non-symmetric, in agreement with our numerical observations.

\begin{figure}[htbp]
\begin{center}
    \begin{equation}
        C_{P_1,P_2} =
            \parbox{0.83\linewidth}{
                \includegraphics[width=\linewidth]{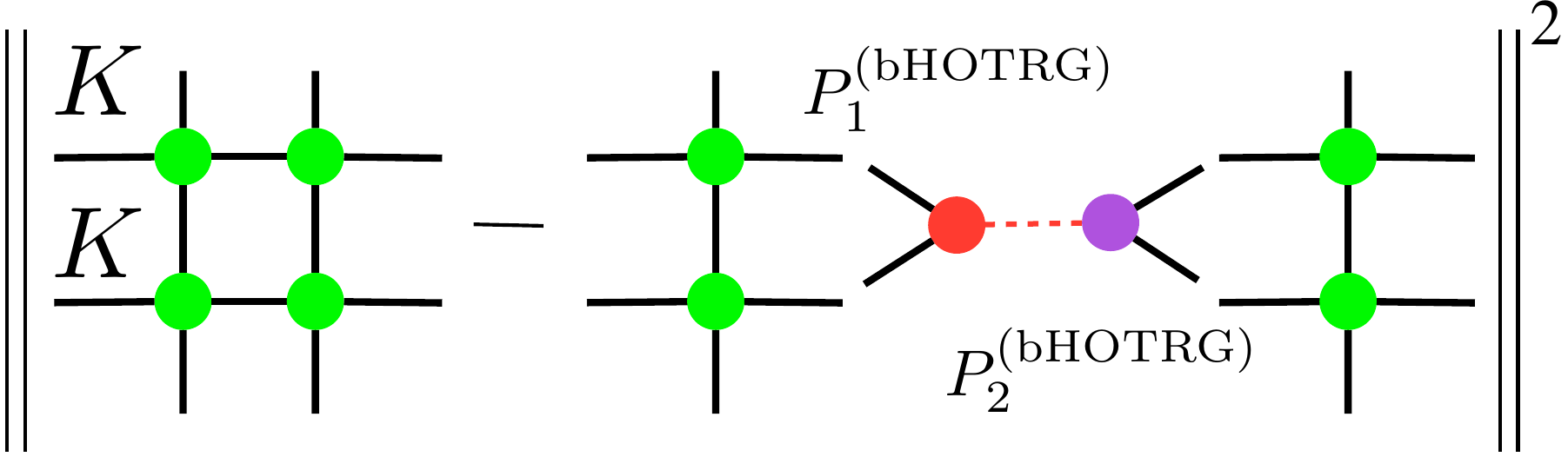}
            }
            \nonumber
    \end{equation}
 \caption{
Cost function of the squeezers $P_1 ^\mathrm{(bHOTRG)}$ and $P_2 ^\mathrm{(bHOTRG)}$ for boundary HOTRG. 
}
\label{fig:Iso_of_bHOTRG}
\end{center}
\end{figure}

In the boundary TRG method this decision is not applied. Instead, squeezers are created from a combination of $U^\mathrm{(HOTRG)}$ and $V^\mathrm{(HOTRG)}$. These squeezers are used for the truncation in the coarse graining step. The procedure minimizes the cost function in \cref{fig:Iso_of_bHOTRG}.
First, the isometries are calculated without truncation as in \cref{eq:Isohotrg} and similarly for $V^\mathrm{(HOTRG)}$. Then, a truncated SVD is performed:
\begin{equation}
    {\lambda ^{(U)}}U^\mathrm{(HOTRG)}V^\mathrm{(HOTRG)}{\lambda ^{(V)}} \simeq U\Lambda V.
\end{equation}

The squeezers can be constructed from these tensors and the previous isometries:
\beqnn{
P_1 ^{\mathrm{(bHOTRG)}}
&\equiv&
V^\mathrm{(HOTRG)}{\lambda ^{(V)}}V^{\dagger}/\sqrt{\Lambda} \\
P_2 ^{\mathrm{(bHOTRG)}}
&\equiv&
(1/\sqrt{\Lambda})
U^\dagger
{\lambda ^{(U)}}
U^\mathrm{(HOTRG)}.
}

The total computational cost is of the same order as the original HOTRG, and the calculation of $P_1$ and $P_2$ is not the dominant cost in the renormalization step.
The results of this boundary HOTRG method are much less dependent on the symmetry properties of the initial tensors as discussed in \cref{sec:InitialTensorDependence}. Therefore, the method creates more reliable results.
In addition, the cost function of the boundary HOTRG in \cref{fig:Iso_of_bHOTRG} approximates four tensors instead of two for the usual HOTRG as in \cref{fig:Iso_of_HOTRG}. The approximation takes into account a larger region and can thus improve the accuracy of the approximation.
We note that the bond-weighted TRG method for HOTRG is also based on the boundary TRG truncation~\cite{bwTRG}.

The ideas presented here can generally be used in any TRG method with isometries.
Replacing $U^\mathrm{(HOTRG)} \rightarrow P_1 ^\mathrm{(bHOTRG)},P_2 ^\mathrm{(bHOTRG)}$ does not require significant additional computational costs but can strongly reduce the initial tensor dependence.

\section{ATRG, MDTRG and variants}
\label{app:ATRG_and_Triad}
We explain the coarse graining steps with ATRG and MDTRG in this appendix. We also introduce variants of the established algorithms and benchmark the different methods for the  two-dimensional Ising model.

The accuracy of the free energy depends on the method used in the coarse-graining step.
Particularly, we observe that algorithms which use isometries to create the indices of the next coarse-grained tensors $K^\mathrm{(next)}$ are highly dependent on the initial tensor properties.

We start from the translationally invariant partition function $Z = \mathrm{tr}\left(\prod_i K_{x_iy_ix' _iy' _i}\right)$, where $x_i$ ($x'_i$) are the indices that connect a lattice point at site $i$ to its nearest neighbor in negative (positive) $x$-direction. Accordingly, $y_i$ ($y'_i$) connects to the next tensor in negative (positive) $y$-direction. Note that $x_{i+1} = x'_{i}$ and $y_{i+1} = y'_{i}$. The trace $\mathrm{tr}$ implies a summation over all indices. $K$ can, for example, be $K^\mathrm{(delta)}$ or $K^\mathrm{(exp)}$ as defined in \cref{sec:InitialTensorDependence}.

Tensor renormalization group algorithms provide a way to coarse-grain a given tensor network to a new network with fewer tensors. This step is approximate to avoid an exponential growth of the numerical costs, and the algorithms differ in the way they truncate the tensors. Typically, two tensors of an initial lattice are replaced by one tensor on a coarse-grained lattice. We restrict ourselves to square lattices in two dimensions but note that most algorithms discussed here can be generalized to higher dimensions. In short, the goal of a tensor renormalization group algorithm is to find the coarse-grained tensor $K^\mathrm{(next)}$ from the initial tensor $K$,
\be{
K_{xyx'y'}
\rightarrow 
K^\mathrm{(next)} _{XYX'Y'}
.
}

\begin{figure}[htbp]
    \begin{center}
         \includegraphics[width=6cm, angle=0]{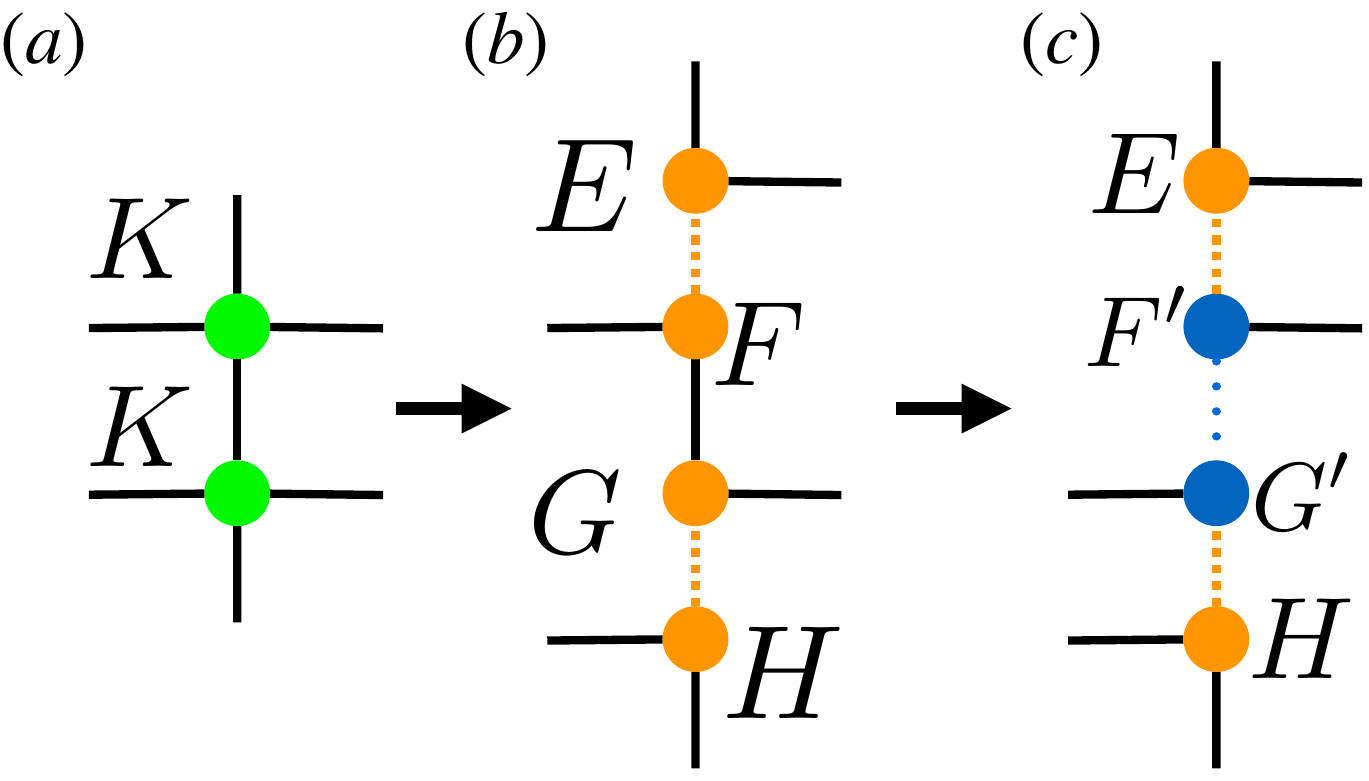}
         \caption{
         Initial steps of a coarse-graining iteration in ATRG-like algorithms. From (a) to (b): the upper (lower) initial tensor is split by an SVD into tensors $E$ and $F$ ($G$ and $H$). From (b) to (c): the tensors $F$ and $G$ are contracted and approximated by an SVD in order to change the index direction. See main text for details.
        }
        \label{fig:ATRG_prepare}
    \end{center}
\end{figure}

For ATRG and MDTRG, we consider two nearest neighbor tensors $\sum_{y}K_{x_1yx '_1y '_1}K_{x_2y_2x '_2y}$ in the coarse-graining step. The tensors are first decomposed into triads, as shown in \cref{fig:ATRG_prepare}(a) to (b). For this, the initial tensors of the translational invariant network are split using an SVD:
\begin{equation}
K_{xyx 'y '}
\simeq
\sum_{b,c} ^{D}
H_{xy b}
\lambda_{bc}
E_{x'y' c}.
\end{equation}
Here, $H$ and $E$ are truncated unitary matrices or isometries, and $\lambda$ is a diagonal matrix with non-negative entries. The smallest singular values are dropped in order not to exceed a maximum bond dimension $D$ in the algorithm. Note that we do not use internal line oversampling in this paper, so we truncate the singular values in intermediate steps to the bond dimension $D$ everywhere.
We define the triad tensors
\begin{align}
    F_{xye} \equiv& \sum_{b}H_{xyb}\lambda_{be} \\
    G_{x'y'g} \equiv& \sum_{c}E_{x'y'c}\lambda_{cg}.
\end{align}
The contraction of two neighboring tensors in the initial network can then be written as
\be{
\sum_{y}K_{x_1yx '_1y '_1}K_{x_2y_2x '_2y}
\simeq
\sum_{y,e,g}
E_{x' _1y' _1e}
F_{x_1ye}
G_{x' _2yg}
H_{x_2y_2g},
}
corresponding to \cref{fig:ATRG_prepare}(b).

\subsection{ATRG and variants}
In the ATRG method, an additional SVD is applied to swap the indices in $x$-direction as shown in \cref{fig:ATRG_prepare}(c):

\begin{align}
    \sum_{y}
    F_{x_1ye}
    G_{x' _2yg}
    \simeq&
        \sum_{f,h} ^{D}
        \Tilde{F'}_{x' _2fe}
        \lambda'_{fh}
        \Tilde{G'}_{x_1hg} \\
    =&
        \sum_{f}
        F' _{x' _2fe}
        G' _{x_1fg}.
\end{align}
The singular values $\lambda'$ are included in $F'$ and $G'$ with square root $\sqrt{\lambda'}$ factors.

\begin{figure}[htbp]
    \begin{center}
    \begin{equation}
        C^\mathrm{ATRG} =
            \parbox{0.83\linewidth}{
                \includegraphics[width=6cm]{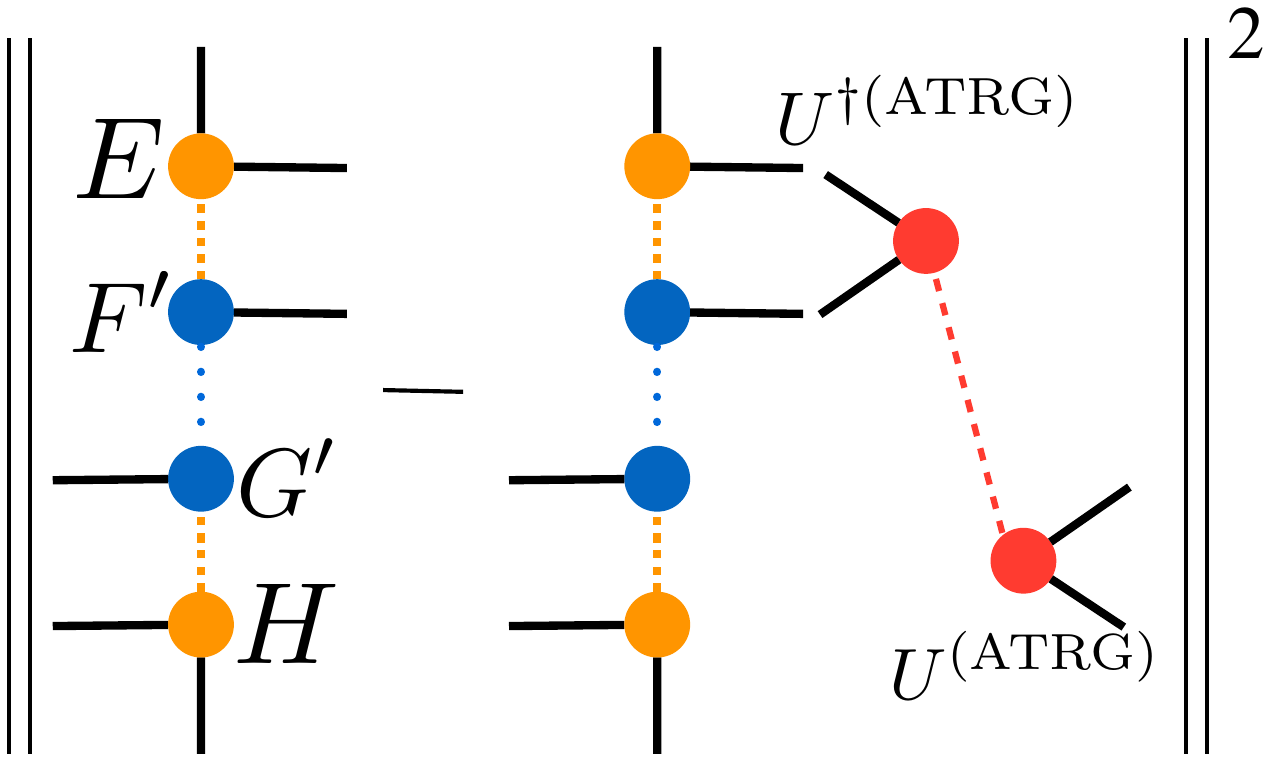}
            }
            \nonumber
    \end{equation}
    \caption{
        Cost function to be minimized by the isometry $U ^\mathrm{(ATRG)}$ in the isometric ATRG algorithm.
    }
\label{fig:Iso_of_ATRG}
\end{center}
\end{figure}

\begin{figure}[htbp]
\begin{center}
 \includegraphics[width=8cm, angle=0]{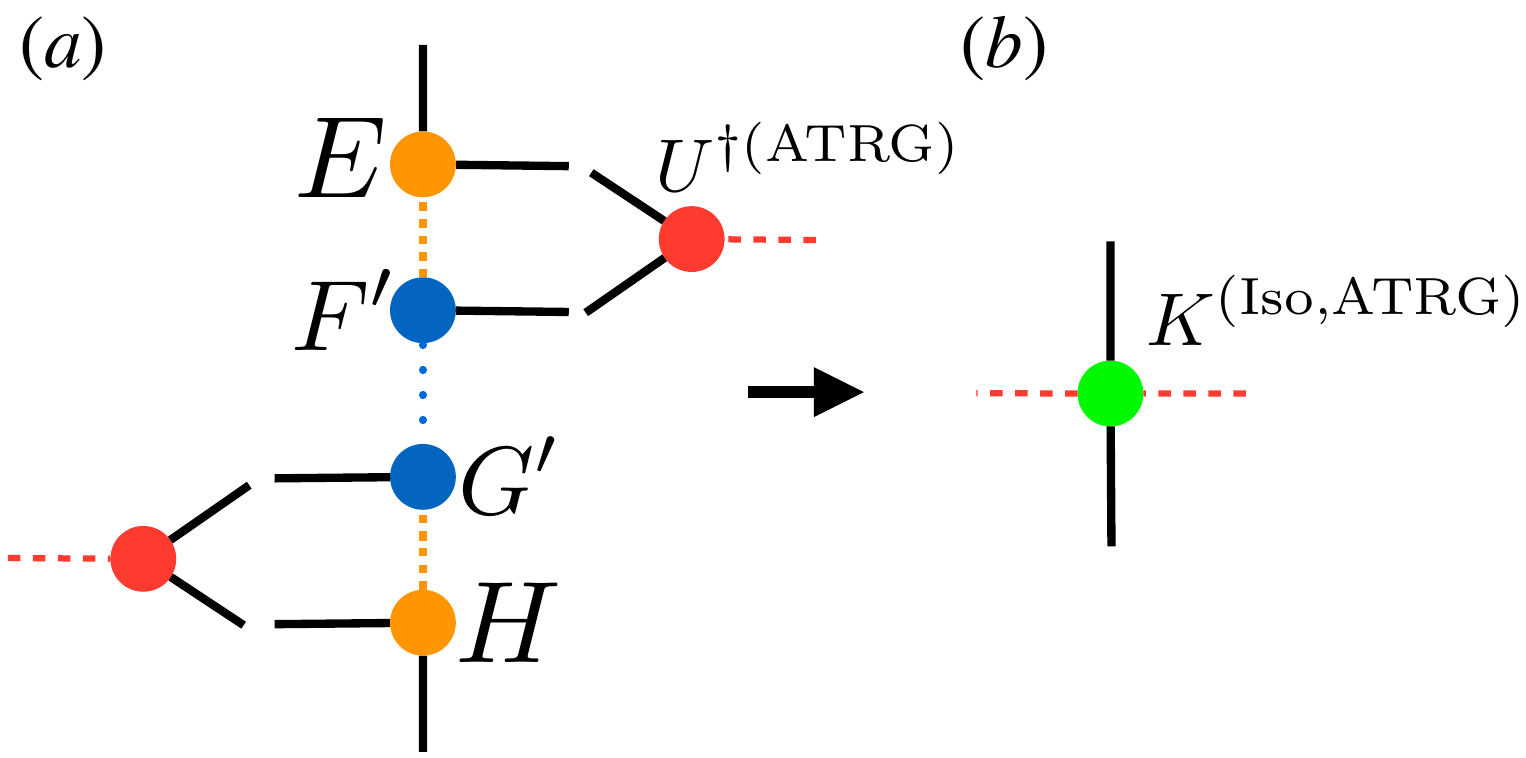}
 \caption{
    Final step to obtain the coarse grained tensor $K^\mathrm{(Iso,ATRG)}$ in the isometric ATRG algorithm. $E,F',G',H$ as in \cref{fig:ATRG_prepare}, $U ^\mathrm{(ATRG)}$ from \cref{fig:Iso_of_ATRG}.
}
\label{fig:IsoATRG_schpic}
\end{center}
\end{figure}

\paragraph{Isometric ATRG.}
In the isometric ATRG, two indices $x_1$ and $x_2$ are combined by applying an isometry $U^\mathrm{(ATRG)}$. This tensor is obtained by an SVD of a combination of triads:
$EF'G'H H ^\dagger {G'} ^\dagger {F'} ^\dagger E ^\dagger  = U^{\mathrm{(ATRG)}} \left(\lambda^\mathrm{(ATRG)}\right)^2 U^\mathrm{\dagger(ATRG)}$.\footref{fn:SVDhermitean}
This minimizes the cost function in \cref{fig:Iso_of_ATRG}.

We finally calculate the coarse-grained tensor $K^{\mathrm{(Iso,ATRG)}}$, as shown in \cref{fig:IsoATRG_schpic},
$K^\mathrm{(Iso,ATRG)} = U^\dagger EF'G'HU$.

\paragraph{ATRG without isometries, and shifted ATRG.}

\begin{figure}[htbp]
\begin{center}
 \includegraphics[width=8cm, angle=0]{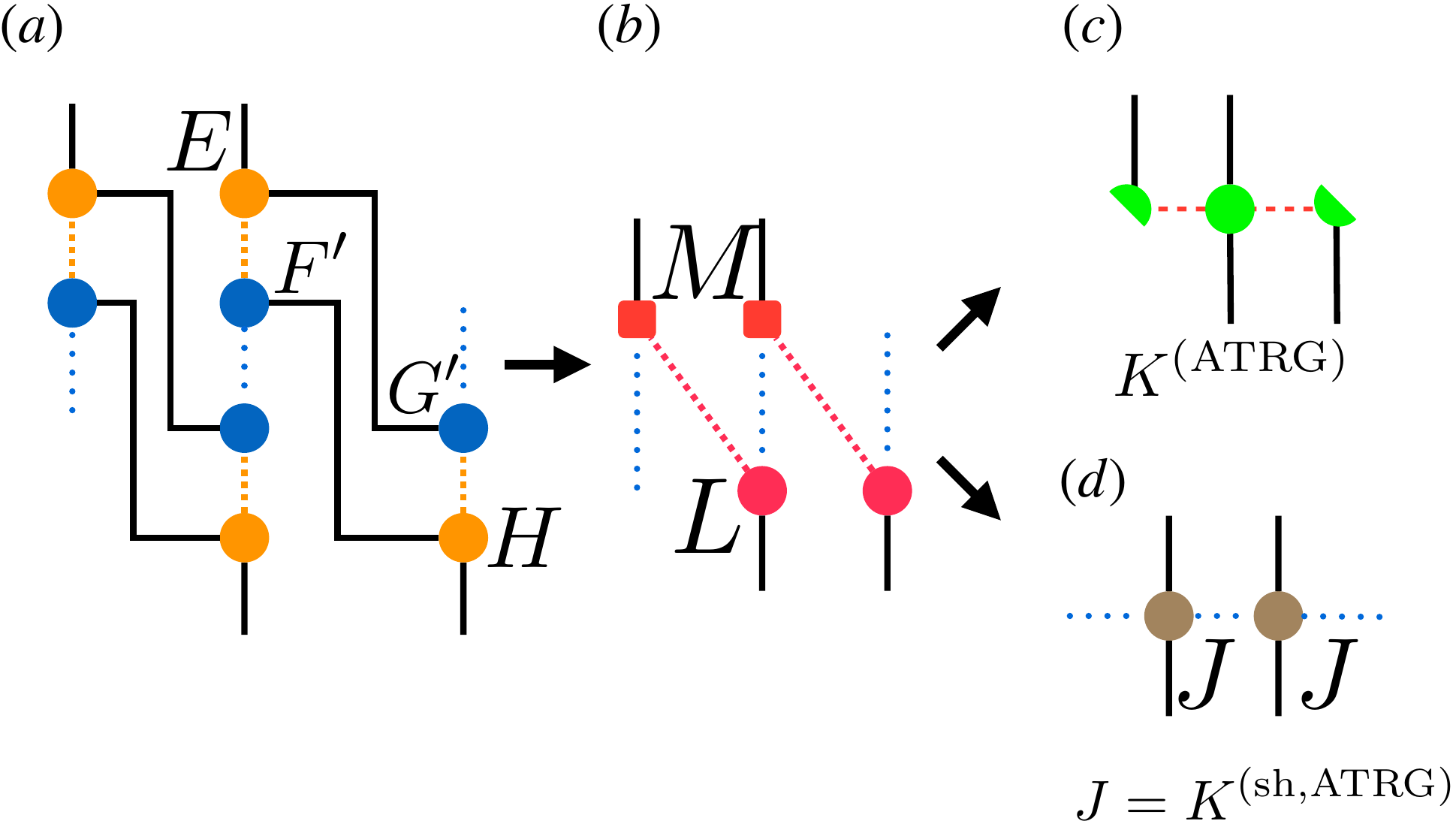}
 \caption{
    Final coarse graining steps of the ATRG and shifted ATRG algorithms. (a) Initial tensor decomposition of four tensors, see \cref{fig:ATRG_prepare} for details. From (a) to (b): the SVD of $EF'G'H$ leads to the tensors $M$ and $L$. From (b) to (c): the tensors $M$ and $L$ are contracted to form the tensor $K^\mathrm{(ATRG)}$ which is the new coarse grained tensor in the ATRG algorithm. From (b) to (d): the tensors $M$ and $L$ are contracted to form the tensor $J$ which is the new coarse grained tensor $K^\mathrm{(sh,ATRG)}$ in the shifted ATRG algorithm. 
}
\label{fig:ATRG_schpic}
\end{center}
\end{figure}

We discuss variants of the ATRG algorithm which do not rely on the applications of isometries as before. Instead, we use further contractions and SVDs. See \cref{fig:ATRG_schpic} for a graphical representation of the individual steps.

First, we take the SVD of the tensor composition $EF'G'H$ from \cref{fig:ATRG_prepare}(b) as

\begin{align}
&\sum_{x'_1,x'_2,e,g}
E_{x'_1 y' e}
F' _{x'_2 f e}
G' _{x'_1 f' g}
H_{x'_2 y g}\nonumber \\
&\simeq
\sum_{X,X'}
\tilde{M}_{fy'X'}
\lambda^{(LM)} _{X'X}
\tilde{L}_{f'yX}
=
\sum_{X}
{M}_{fy'X}
{L}_{f'yX}\nonumber \\
&=
J_{fyf'y'}.
\label{eq:shATRG_J}
\end{align}

We define the shifted ATRG, which takes these tensors $J$ as the coarse grained tensors:
\be{
K^{\mathrm{(sh,ATRG)}} _{XyX'y'}
    =
J_{XyX'y'}.
\label{eq:shATRG_K}
}
Alternatively, another contraction defines the coarse grained tensor of ATRG without isometry,
\begin{equation}
    K^{\mathrm{(ATRG)}} _{XyX'y'}
    =
    \sum_{f}
    M_{fy'X'}
    L_{fyX}.
    \label{eq:Katrg}
\end{equation}

The SVD which leads to $M$ and $L$ requires $\order(D^6)$ operations if we do not apply a truncated SVD method. If we apply the ideas of the randomized SVD instead, the costs can be reduced to $\order(D^5)$. See~\cite{RandTRG,TriadRG,MDTRG} for more details.

The method to create the coarse-grained tensors $K^{\mathrm{(ATRG)}}$ is equivalent to the original introduction of ATRG in~\cite{ATRG}. 
The original ATRG method can be understood as a replacement of the isometries $U^{(\mathrm{ATRG})}$ in the isometric ATRG as in \cref{fig:IsoATRG_schpic} by squeezers. These originate from the truncated SVD in \cref{eq:shATRG_J}. Explicitly, the squeezers are:
\begin{align}
    P_1^{(\mathrm{ATRG})}
    =&
    [G'H]
    \tilde{L}^\dagger/\sqrt{\lambda ^{(LM)}} \\
    P_2^{(\mathrm{ATRG})}
    =&
    (1/\sqrt{\lambda ^{(LM)}})
    \tilde{M}^\dagger
    [EF'].
\end{align}

The algorithms which create $K^{\mathrm{(ATRG)}}$ and $K^{\mathrm{(Iso,ATRG)}}$ differ in the regions that are approximated in the truncation step, and in the way how the coarse-grained tensors are constructed. 
The shifted ATRG also creates a different approximation compared to $K^\mathrm{(ATRG)}$. This method can, however, only be used to coarse-grain the indices in one direction. For example, the tensor $J$ would have additional indices for the $z$-direction in three dimensions. $J_{Xyz_1z_2X'y'z_1 'z_2 '}$ has $D^8$ elements, and creating it directly is not possible within the leading costs of $\order(D^7)$ for the ATRG methods in three dimensions. Shifted ATRG is thus only applicable for two-dimensional systems or in combination with other methods which coarse-grain the additional directions beforehand.
The other two ATRG methods (isometric ATRG and ATRG) can be directly generalized to higher dimensions~\cite{ATRG}.

\paragraph{Shifted isometric ATRG.}

\begin{figure}[htbp]
\begin{center}
 \includegraphics[width=8cm, angle=0]{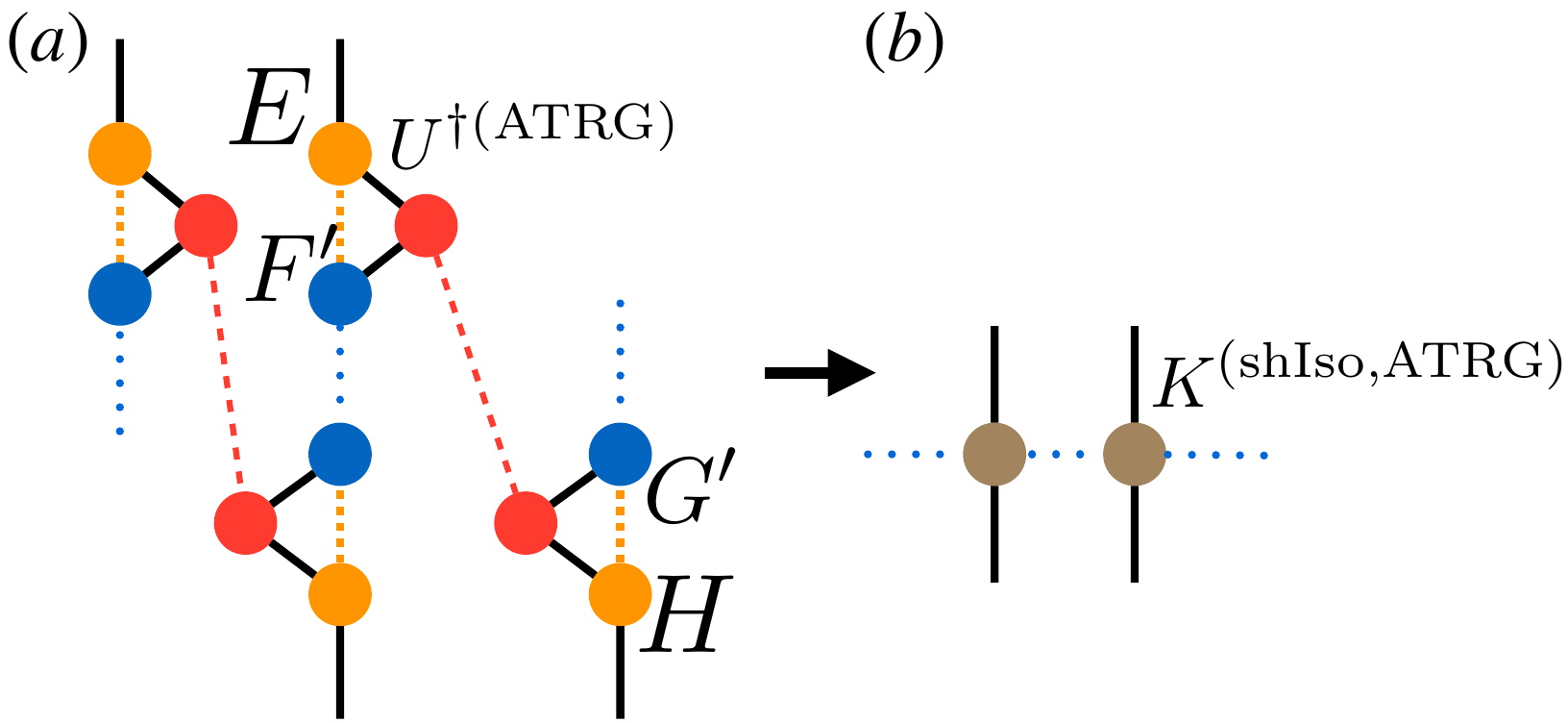}
 \caption{
    Final coarse-graining steps of the shifted isometric ATRG. (a) The initial tensor decomposition of four tensors (see \cref{fig:ATRG_prepare} for details) is approximated by inserting the isometries from \cref{fig:Iso_of_ATRG}. (b) This leads to the new coarse-grained tensors $K^\mathrm{(shIso,ATRG)}$.
}
\label{fig:shIsoATRG_schpic}
\end{center}
\end{figure}

Instead of using randomized techniques for the contraction or SVD of $J$, we can approximate the contraction using the isometry $U^\mathrm{(ATRG)}$ that was introduced for the isometric ATRG: $J = EF'G'H\simeq EF'U^\mathrm{(ATRG)}U^\mathrm{\dagger (ATRG)} G'H$. We call this method shifted isometric ATRG. It is shown in \cref{fig:shIsoATRG_schpic}. Note that the isometry does not create the indices of the coarse-grained tensors directly, since all indices of $U^\mathrm{(ATRG)}$ are contracted.
This approximation of $J$ may not be optimal, because the isometry is not calculated from the same subregion of the tensor network as $J$ itself: $U$ optimizes $(HG')(F'E)UU^\dagger$, not $(F'E)UU^\dagger(HG')$. We include this method in our benchmark, however, to test the accuracy of a method that uses isometries for the contractions.

\begin{figure}[htbp]
\begin{center}
    \begin{equation}
        C^\mathrm{MDTRG} =
            \parbox{0.83\linewidth}{
                \includegraphics[width=6cm]{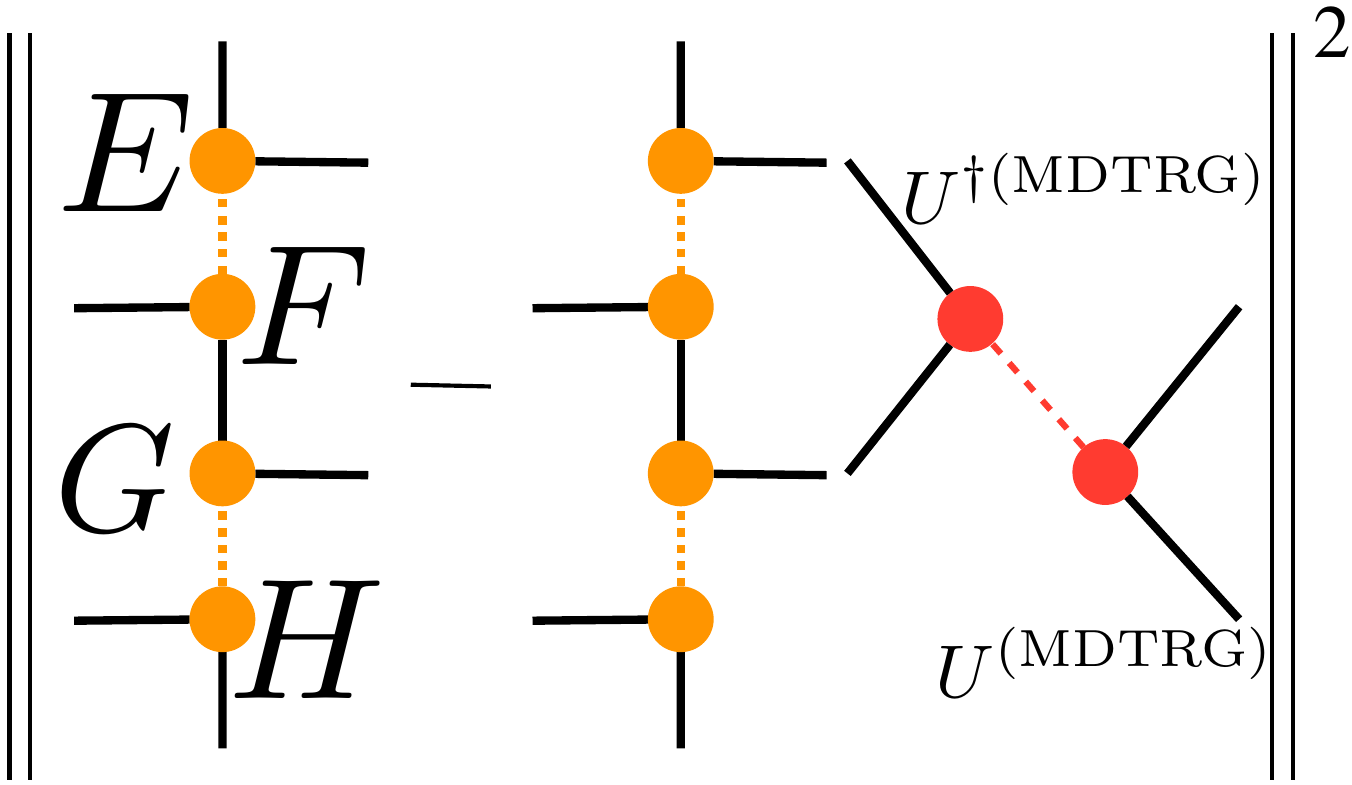}
            }
            \nonumber
    \end{equation}
 \caption{
    Cost function of the isometry $U ^\mathrm{(MDTRG)}$ for MDTRG.
}
\label{fig:Iso_of_MDTRG}
\end{center}
\end{figure}

\begin{figure}[htbp]
\begin{center}
 \includegraphics[width=8cm, angle=0]{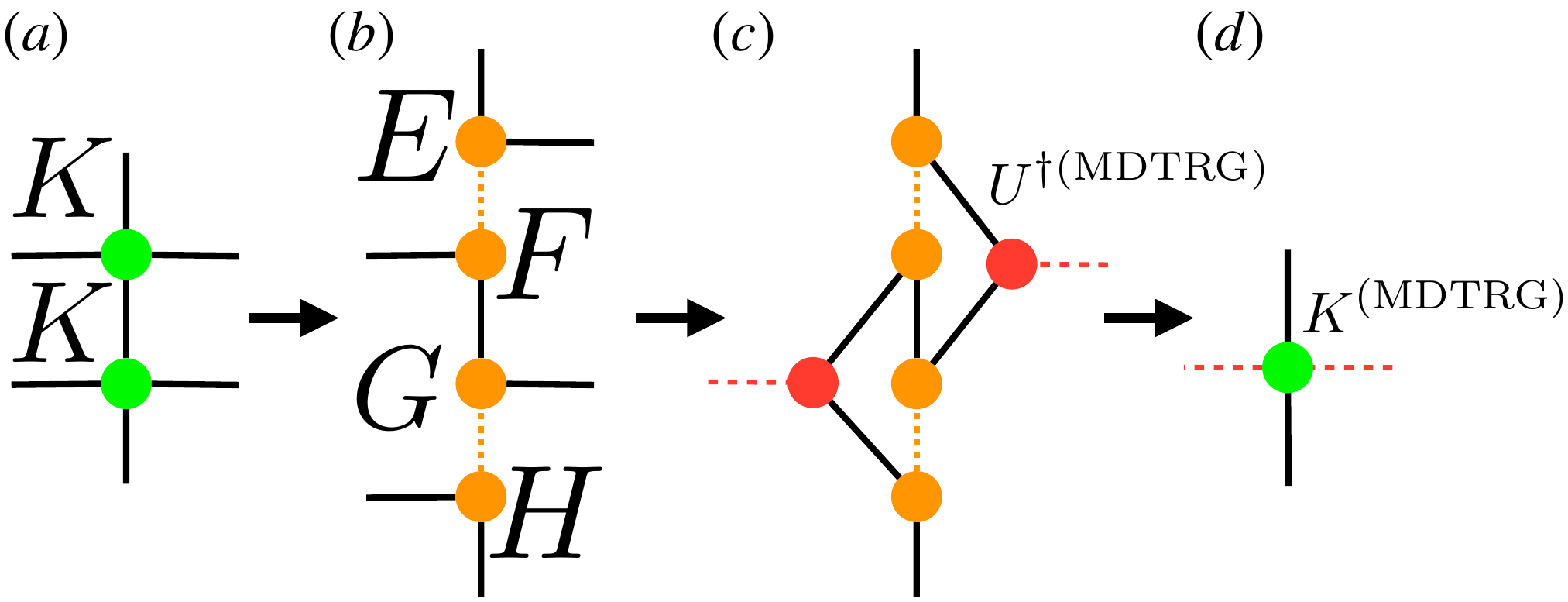}
 \caption{
         Coarse-graining iteration of the MDTRG algorithm. From (a) to (b): the upper (lower) initial tensor is split by an SVD into tensors $E$ and $F$ ($G$ and $H$), similar to ATRG (see \cref{fig:ATRG_prepare}). From (b) to (c): isometries are applied, which minimize the cost function in \cref{fig:Iso_of_MDTRG}. From (c) to (d): an approximate contraction leads to the coarse-grained tensors $K^\mathrm{(MDTRG)}$.
}
\label{fig:MDTRG_schpic}
\end{center}
\end{figure}

\subsection{MDTRG and variants}
In the following we explain the MDTRG method and also introduce a variation of it. 
The method is similar to the TTRG~\cite{TriadRG} but with a different approximation in the contraction step.
Compared to ATRG, the index swapping (from (b) to (c) in \cref{fig:ATRG_prepare}) is omitted and the tensors $EFGH$ are directly used instead of $EF'G'H$.

For MDTRG, we calculate the isometry $U^\mathrm{(MDTRG)}$ from the tensors $EFGH$. The cost function is shown in \cref{fig:Iso_of_MDTRG}. Namely, we use the decomposition $EFGHH^\dagger G^\dagger F^\dagger E^\dagger =U^\mathrm{(MDTRG)}\left(\lambda^\mathrm{(MDTRG)}\right)^2U^\mathrm{\dagger(MDTRG)}$. Note that the lefthand-side of this equation is Hermitian, and thus the left- and right-singular vectors are equal on the righthand-side.\footref{fn:SVDhermitean} The coarse grained tensor is then obtained, as shown in \cref{fig:MDTRG_schpic}, by a contraction with the isometries:
\begin{align}
    K^{\mathrm{(MDTRG)}}
    _{XyX'y'}
    =&
    \sum_{x_1,x_2,x'_1,x'_2,y,e,g}
    U_{x'_1 x'_2 X'} ^\mathrm{*MDTRG)}
    U_{x_1 x_2 X} ^\mathrm{(MDTRG)}\nonumber\\
    &\times
    E_{x'_1 y' e}
    F_{x_1 y e}
    G_{x'_2 y g}
    H_{x_2 y g}.
\label{mdtrgcont}
\end{align}
This contraction requires a truncated SVD method to reduce the costs, in two dimensions to $\order(D^5)$. We use the randomized SVD, as in~\cite{MDTRG}.

Using the approximation, we get the triad representation of the $K^\mathrm{(MDTRG)}$ as the SVD of $K^{\mathrm{(MDTRG)}}$ with square root weight,
\be{
K^{\mathrm{(MDTRG)}}
_{AyA'y'}
\simeq
\sum_{n}
N_{Ayn}
O_{A'y'n}.
}

Replacing the isometries by squeezers and applying the ideas of the boundary TRG (see \cref{app:b_TRG}) to MDTRG is straightforward. For this boundary MDTRG method, we calculate $U^\mathrm{(MDTRG)}$ and $V^\mathrm{(MDTRG)}$ by a randomized SVD with oversampling size $rD$, and compute the squeezers $P_1 ^\mathrm{(MDTRG)}$ and $P_2 ^\mathrm{(MDTRG)}$ from this.

\begin{figure}[htbp]
\begin{center}
 \includegraphics[width=8cm, angle=0]{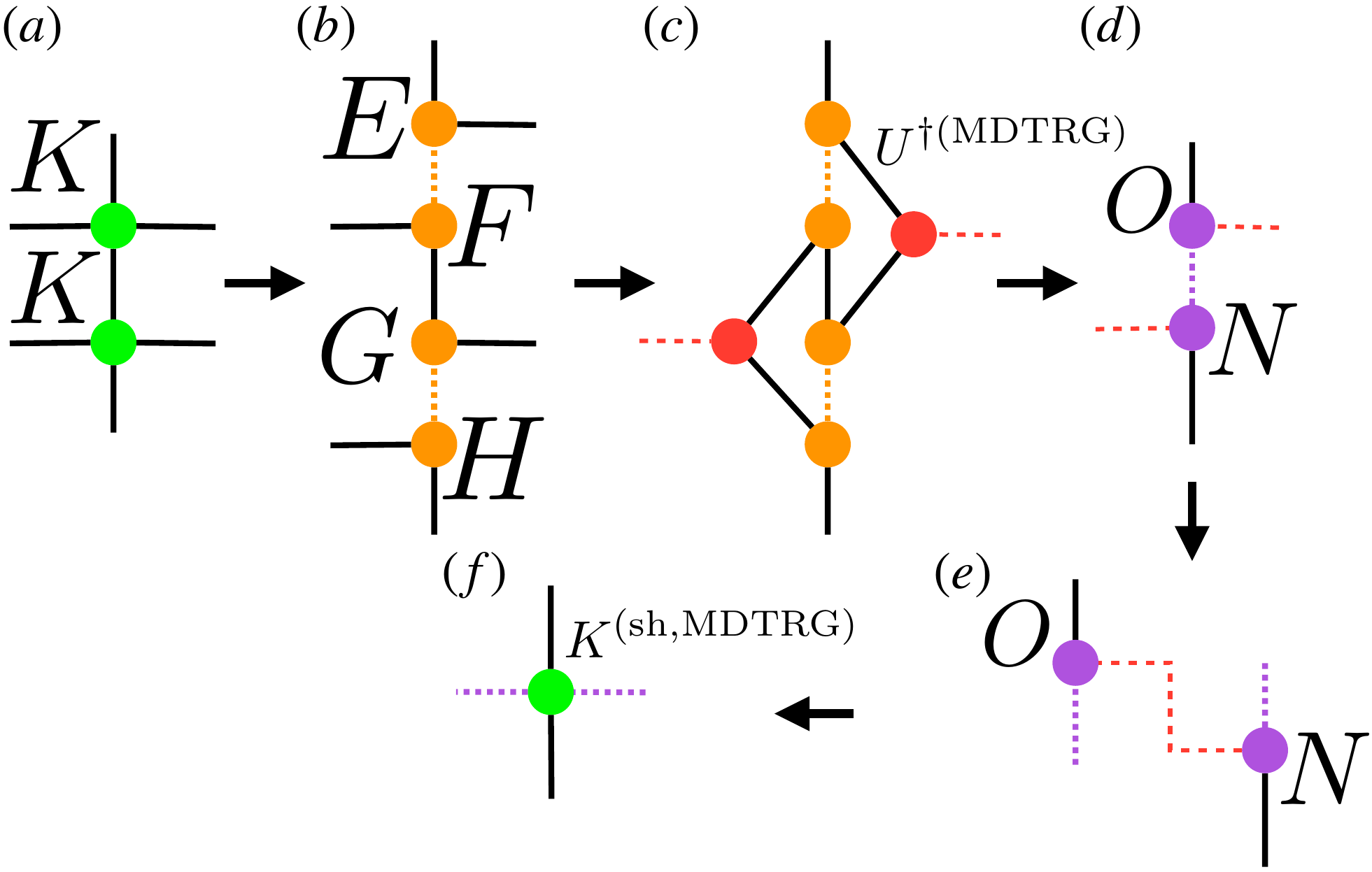}
 \caption{
         Coarse-graining iteration of the MDTRG algorithm. From (a) to (b): the upper (lower) initial tensor is split by an SVD into tensors $E$ and $F$ ($G$ and $H$), similar to ATRG (see \cref{fig:ATRG_prepare}). From (b) to (c): isometries are constructed, which minimize the cost function in \cref{fig:Iso_of_MDTRG}. From (c) to (d): an approximate SVD of the tensor network in (c) leads to the the decomposition into $O$ and $N$, with square root weighted singular values included. From (d) to (e): the indices are shifted, making use of translational invariance. From (e) to (f): contraction of $O$ and $N$ leads to the coarse-grained tensors $K^\mathrm{(sh,MDTRG)}$.
}
\label{fig:shMDTRG_schpic}
\end{center}
\end{figure}

Furthermore, we define the shifted MDTRG as depicted in \cref{fig:shMDTRG_schpic}. In the previous MDTRG algorithm, the tensors $EFGH$ and the isometries were contracted to form the new coarse-grained tensors. Instead, the shifted MDTRG replaces this contraction by an approximate SVD, which can be applied efficiently to the tensor network.
From this tensor decomposition we obtain truncated unitaries, which are combined with the square roots of the singular values to form new tensors $O$ and $N$.
Their contraction leads to the coarse-grained tensors:
\be{
K^{\mathrm{(sh,MDTRG)}}
_{XyX'y'}
\simeq
\sum_{A}
N_{AyX'}
O_{Ay'X}.
} 
Note that the index that was created in the SVD forms one of the indices of the coarse grained tensor. 
\Cref{fig:shMDTRG_schpic} shows the contraction for shifted MDTRG as (e) to (f).
New indices of the shifted MDTRG are dotted purple line which comes from the truncated SVD of \cref{mdtrgcont}.
\subsection{Comparison of coarse-graining methods}

We benchmark the different TRG algorithms for the two-dimensional critical Ising model. The results are summarized in \cref{tab:ATRGMDTRG} and discussed in the main text. As mentioned there, we divide the algorithm into three classes. Algorithms denoted as \textit{iso} in \cref{tab:ATRGMDTRG} apply isometries to the tensors to create the coarse grained indices. The methods marked as \textit{iso$*$} use isometries as well, but only for intermediate contraction steps, and the final coarse-grained indices are not directly the truncated indices of the isometries. Finally, all other algorithms are marked as \textit{sqz}.

When isometries are introduced in a tensor network to combine bonds and to compress the bond dimension, there is an ambiguity in choosing these tensors. They can be optimized for either direction of the bond that shall be compressed. An example can be seen in \cref{fig:Iso_of_HOTRG}, where the isometries $U$ and $V$ minimize the error with respect to different contraction directions. Only one of the two is chosen in isometric algorithms for the coarse-graining, and this can lead to a significant decrease in the accuracy if the tensor is not symmetric. Otherwise, $U$ and $V$ are identical and the problem does not arise. The squeezers introduced in the boundary TRG algorithm and discussed in \cref{app:b_TRG} take into account both isometries. Thus, these methods do not suffer from the errors introduced by omitting the other isometry.

Even though the original TRG algorithm uses an SVD as well for the coarse-graining, both isometries are used in this case. This makes it equivalent to the squeezer algorithms and we group it as \textit{sqz}.

\begin{figure}[htbp]
\begin{center}
 \includegraphics[width=8cm, angle=0]{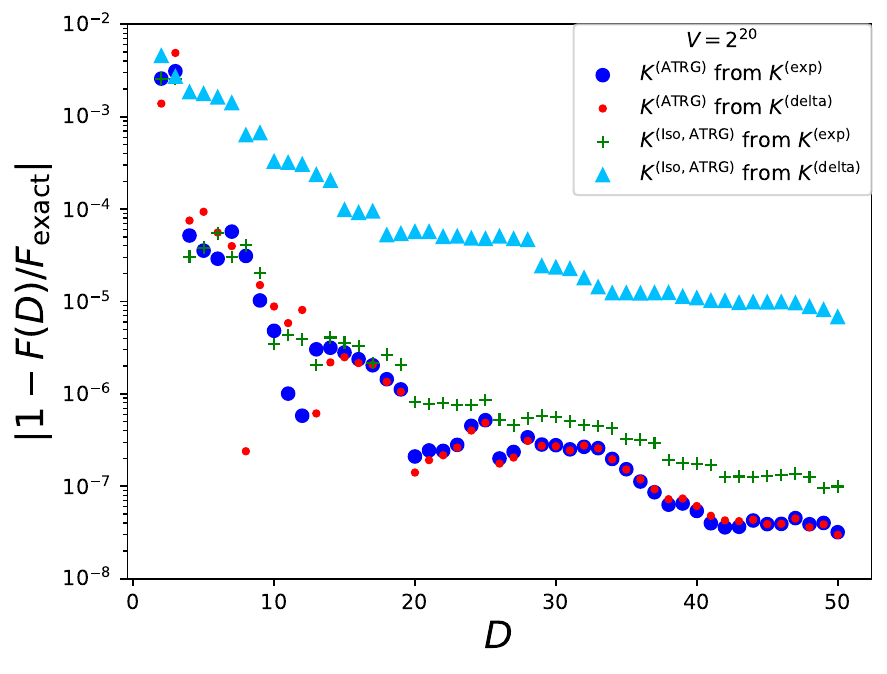}
 \caption{
Relative error of the free energy of the two-dimensional Ising model at the critical temperature for different bond dimensions $D$. Comparison between ATRG and isometric ATRG for the symmetric initial tensor $K^\mathrm{(exp)}$ and the non-symmetric initial tensor $K^\mathrm{(delta)}$.
}
\label{Appfig:ATRG}
\end{center}
\end{figure}

\begin{figure}[htbp]
\begin{center}
 \includegraphics[width=8cm, angle=0]{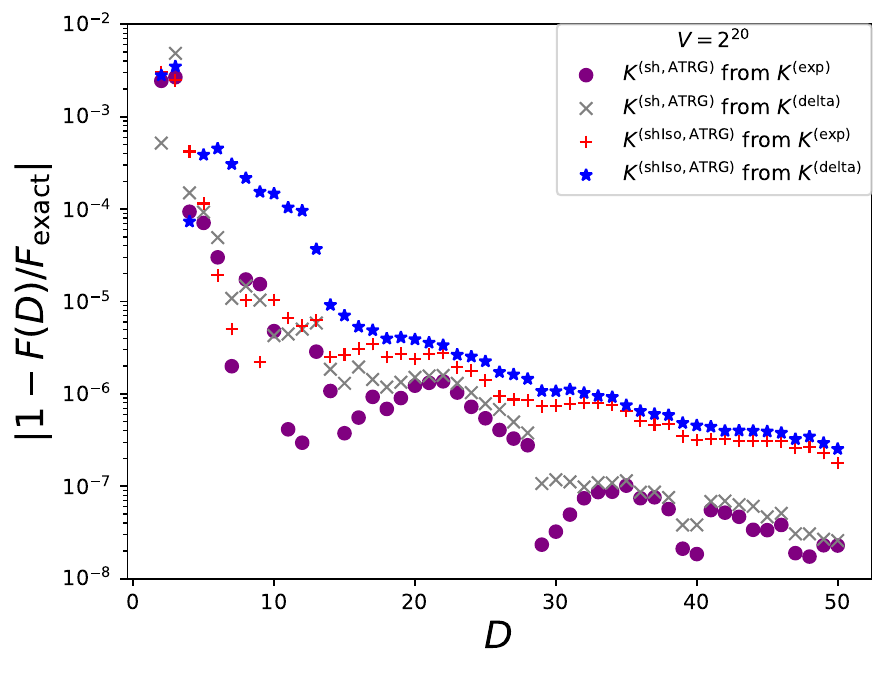}
 \caption{
    Relative error of the free energy of the two-dimensional Ising model at the critical temperature for different bond dimensions $D$. Comparison between shifted ATRG and shifted isometric ATRG for the symmetric initial tensor $K^\mathrm{(exp)}$ and the non-symmetric initial tensor $K^\mathrm{(delta)}$.
}
\label{Appfig:shATRG}
\end{center}
\end{figure}

\begin{figure}[htbp]
\begin{center}
 \includegraphics[width=8cm, angle=0]{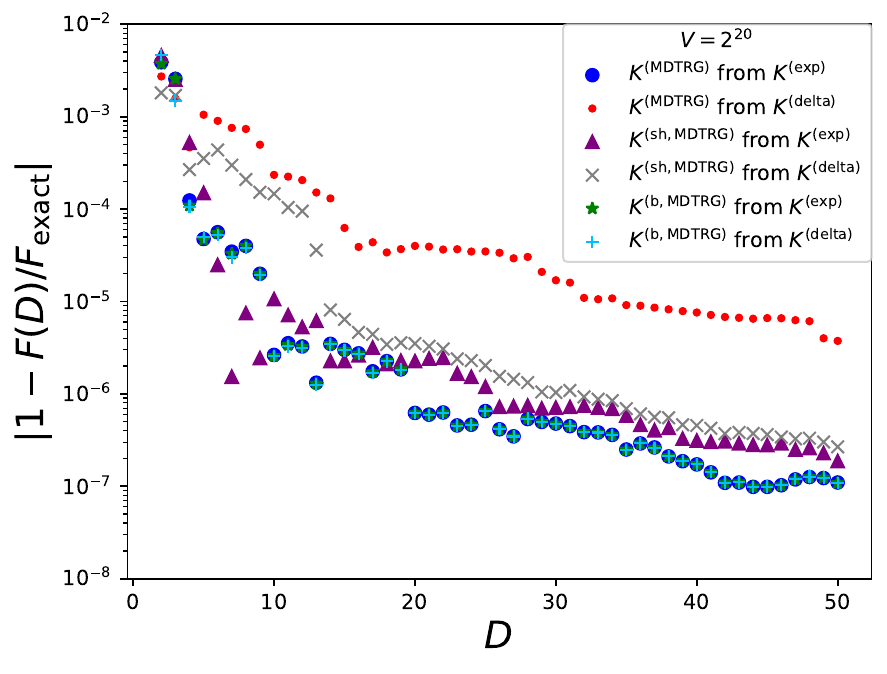}
 \caption{
    Relative error of the free energy of the two-dimensional Ising model at the critical temperature for different bond dimensions $D$. Comparison between MDTRG, shifted MDTRG and boundary MDTRG for the symmetric initial tensor $K^\mathrm{(exp)}$ and the non-symmetric initial tensor $K^\mathrm{(delta)}$.
}
\label{Appfig:MDTRG}
\end{center}
\end{figure}

Our benchmark results for the ATRG and MDTRG methods are shown in \cref{Appfig:ATRG,Appfig:shATRG,Appfig:MDTRG}. For the truncated SVD, we use the randomized SVD with an oversampling parameter $r=4$, such that the SVD is performed in an $rD$ dimensional subspace. We test all methods with two initial tensors, a symmetric tensor $K^{\mathrm{(exp)}}$ (see \cref{eq:2d_ising_Kexp}) and a non-symmetric one $K^{\mathrm{(delta)}}$ from our initial tensor construction (see \cref{eq:2d_ising_Kdelta}).

\Cref{Appfig:ATRG} shows that the ATRG does not only produce more accurate results for large bond dimensions compared to the isometric ATRG. Also, ATRG (type \textit{sqz}) shows no dependence on the initial tensors, while isometric ATRG (type \textit{iso}) has a strong dependence and is much less accurate for the non-symmetric initial tensor.

Both shifted ATRG methods (shifted ATRG, type \textit{sqz} and isometric shifted ATRG, type \textit{iso$*$}) show only a very mild dependence on the initial tensors as can be seen in \cref{Appfig:shATRG}. The shifted ATRG has a similar accuracy compared to the common ATRG. Combined with the technical advantages discussed in \cref{app:imp_ATRG}, this method makes a good candidate for the impurity tensor method to calculate observables.

For the MDTRG methods shown in \cref{Appfig:MDTRG}, we find that the MDTRG (type \textit{iso}) produces much less accurate results if a non-symmetric initial tensor is chosen. If the boundary TRG method is applied (type \textit{sqz}), the results coincide with those of the usual MDTRG method and symmetric initial tensors. The boundary MDTRG obtains similar results, however, for non-symmetric tensors as well. This shows again how the squeezers can make observables more resilient against the choice of initial tensors. The shifted MDTRG (type \textit{iso$*$}) shows only a mild dependence on the initial tensors but has slightly larger errors than boundary MDTRG for large bond dimensions.

From our numerical calculations with the variants of the ATRG and MDTRG, TRG, and HOTRG, we find that the TRG methods with coarse-grained tensors $K^\mathrm{(next)}$, whose indices are directly created from isometries, have large initial tensor dependencies.
This dependence is eliminated if we apply the boundary TRG technique as discussed in \cref{app:b_TRG}. We therefore recommend the truncation method with squeezers based on the boundary TRG method, which does not increase the numerical costs significantly but leads to more reliable results.

%
%
%
%
%
%
%
%
%
\section{Impurity tensor method for ATRG}
\label{app:imp_ATRG}
Impurity tensors can be used to calculate physical observables with TRG methods. We give a brief introduction and overview and discuss the differences that arise for the ATRG and the shifted ATRG method. The latter was introduced in \cref{app:ATRG_and_Triad}.
The impurity tensor method was first suggested in~\cite{PhysRevB.78.205116}. It is elsewhere discussed in much detail for TRG~\cite{impurity} and also for HOTRG (with isometries)~\cite{Morita:2018tpw}.
The method is widely used for TRG calculations, see e.g.~\cite{PhysRevD.100.054510,Jha_2020,PhysRevB.103.245137,PhysRevD.104.054505,RevModPhys.94.025005,Samlodia:2024kyi}.

In tensor renormalization group methods, the partition function $Z$ is represented by a translational invariant repetition of a tensor $T_{abcd}(\beta)$ in a volume $V$ as 
\be{
    Z
    =
    \mathrm{tr}
    \prod_{i=1} ^V
    T_{a_ib_ic_id_i}(\beta).
}
We assume the tensor $T(\beta)$ is a function of a parameter $\beta$, which could, for example, be the inverse temperature. Using the product rule and exploiting the translational invariance of the network, the derivative of Z with respect to $\beta$ is
\be{
    \frac{1}{V}\frac{\partial Z}{\partial \beta}
        =
    \mathrm{tr}        \left(\frac{\partial T_{a_1b_1c_1d_1}}{\partial \beta}\right)    \prod_{i=2} ^V
   T_{a_ib_ic_id_i}(\beta).
}
We call $\left(\frac{\partial T}{\partial \beta}\right)$ the impurity tensor.

\begin{figure}[htbp]
    \begin{center}
         \includegraphics[width=6cm, angle=0]{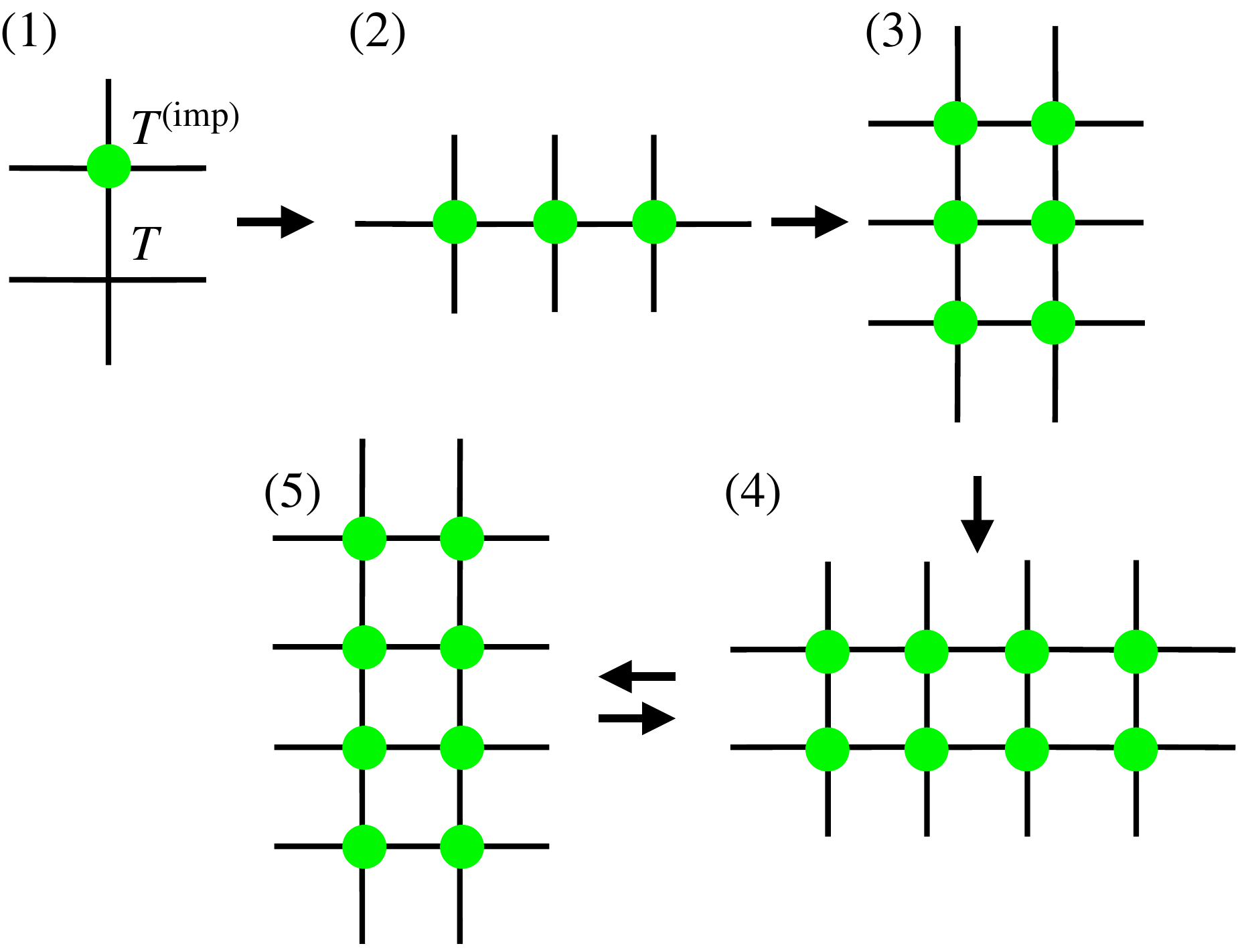}
         \caption{
            Propagation of the impurity tensor information for ATRG. The green dots represent tensors which include information of the impurity and differ from the other tensors. The spread of information can be understood from \cref{fig:ATRG_schpic}, going from (b) to (c). The information of a pair of tensors in the vertical direction (see \cref{fig:ATRG_prepare}) is spread to three coarse-grained tensors in the horizontal direction in \cref{fig:ATRG_schpic} (c). We refer to the vertical and horizontal directions as shown in \cref{fig:ATRG_schpic}. In the figure above, the lattice orientation does not change. From (1) to (2): The first coarse-graining step leads to three affected tensors. From (2) to (3): After exchanging the $x-$ and $y-$direction, two pairs in the vertical direction are affected, and each pair creates three tensors with impurity information. The total number of impurity tensors is thus six. (4) and (5): Following the same logic, the information ultimately spreads to eight tensors for the ATRG algorithm.
        \label{fig:ATRG_imp}
        }
    \end{center}
\end{figure}

\begin{figure}[htbp]
    \begin{center}
         \includegraphics[width=6cm, angle=0]{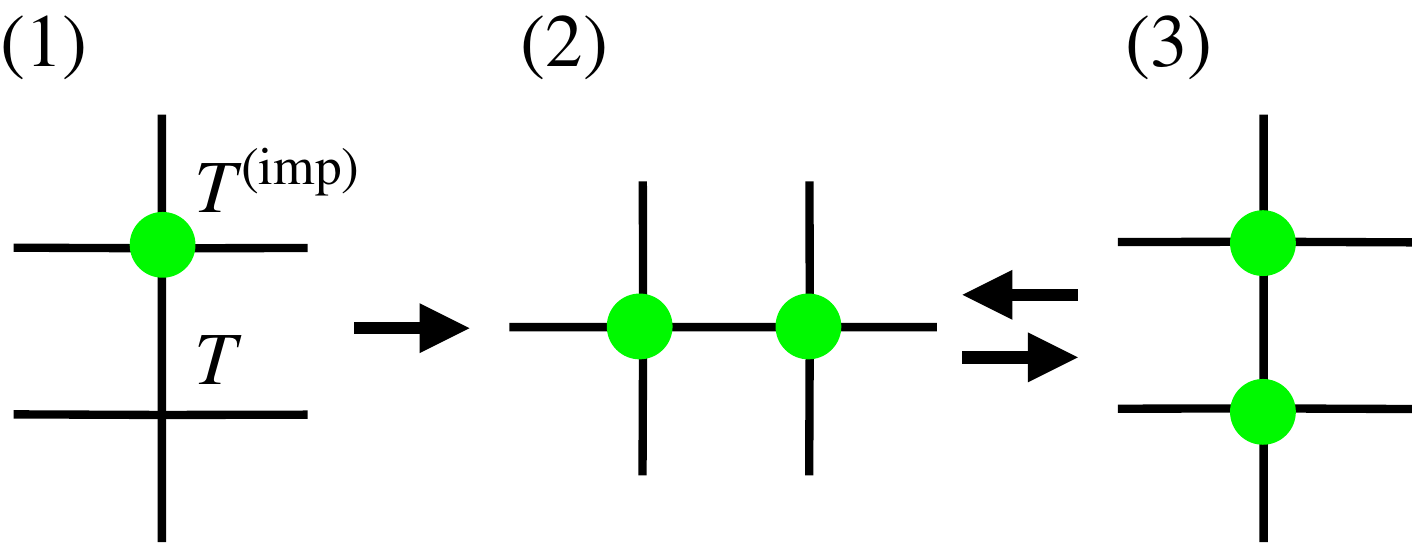}
         \caption{
            Propagation of the impurity tensor information for shifted ATRG. The green dots represent tensors which include information of the impurity and differ from the other tensors. From (1) to (2): Initially, the information is spread to two tensors. From (2) to (3): In \cref{fig:ATRG_schpic}, the impurity information is contained in the four central tensors in step (b). These are contracted to two coarse-grained tensors in (d), which leads to two impurity tensors after the coarse-graining. From (2) to (3): In the next steps, an exchange of $x-$ and $y-$directions leads to a similar picture, where the information is again only contained in the four central tensors in (b). These are contracted to two coarse-grained impurity tensors. This keeps on repeating in subsequent steps, and only two impurity tensors have to be kept with the shifted ATRG.
        }
        \label{fig:ATRG_imp_2}
    \end{center}
\end{figure}

In the impurity tensor method, we need to consider the propagation of the impurity tensor information in each coarse graining step.
In order to keep the information of the impurity tensor at each step, we have to store sub tensor networks~\cite{impurity}.
For the simple TRG, we need to store four different tensors.
We show how ATRG (\cref{fig:ATRG_imp}) and its variation shifted ATRG (\cref{fig:ATRG_imp_2}) can be used for the impurity tensor method.

With the original ATRG, the information of the initial impurity tensor propagates to eight different tensors in later coarse-graining steps, as is shown in \cref{fig:ATRG_imp}.
In contrast to this we only need to calculate and store two coarse-grained impurity tensors with the shifted ATRG, as is shown in \cref{fig:ATRG_imp_2}. The difference arises from the contraction step in \cref{fig:ATRG_schpic}. There, the tensor network $EFGH$ contributes to three coarse-grained tensors $K^\mathrm{(ATRG)}$ in original ATRG (from \cref{fig:ATRG_schpic}(b) to (c)). In contrast to this, the tensors $EFGH$ only affect two coarse-grained tensors $K^\mathrm{(sh,ATRG)}$ for the shifted ATRG, see \cref{fig:ATRG_schpic}(b) to (d).
We use the shifted ATRG method to calculate the free energy in the $\mathbb{Z}_2$ gauge theory (see \cref{sec:Z2}) because of the lower memory footprint and computational costs.

In our initial tensor construction, the original, not locally connected tensor formulation of the partition function or any observable can be site dependent. For example, an observable could be expressed by a product of tensors in \cref{eq:Ising2d_nonlocal}, but potentially with site dependent tensors $K(x,y)$. This does not affect the construction of the locally connected tensor network by delta functions in the subsequent steps. Indeed, translational invariance is not required for our initial tensor construction. Only periodic boundary conditions are assumed instead in order to shift the indices of the delta matrices, for example from \cref{eq:Ising2d_nonlocal} to \cref{eq:TN2dim}. While common TRG algorithms assume translationally invariant systems, impurity tensor methods can be applied to coarse-grain the resulting tensor network with site-dependent tensors. This allows, for example, for the calculation of first and higher-order derivatives and correlation functions. Potentially, several impurity tensors need to be included in the coarse-graining steps.
%
%
%
%
%
%
%
%
%
%

\section{Index direction swapping}
\label{app:xyrot}

\begin{figure}[htbp]
\begin{center}
 \includegraphics[width=8cm, angle=0]{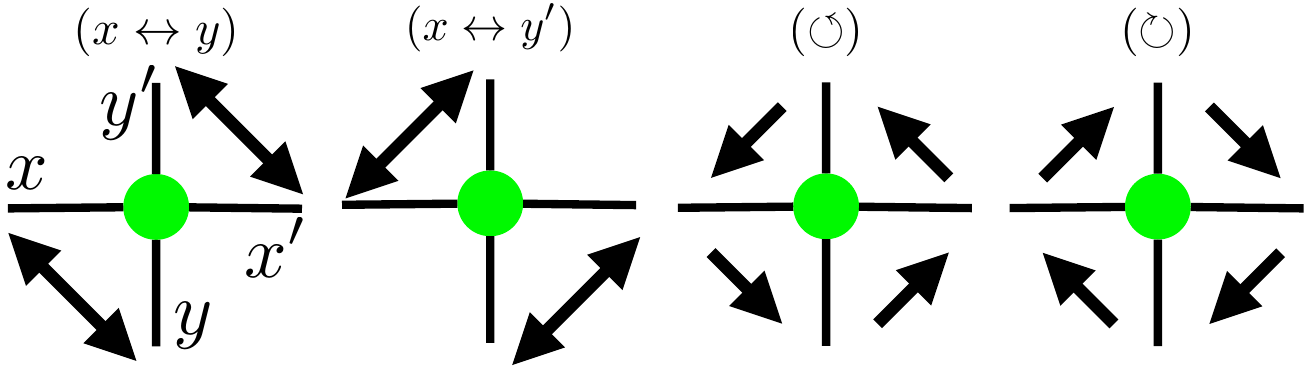}
 \caption{
Exchange of indices after each coarse-graining step for a two-dimensional system: $xy$ flips and rotations. 
}
\label{Appfig:rotExplanation}
\end{center}
\end{figure}

In a TRG coarse-graining step, two initial tensors $K$ are combined into a single new tensor $K^\mathrm{(next)}$. This was explained in \cref{app:ATRG_and_Triad} for two tensors connected by a link in $y$-direction.
For a two-dimensional lattice, this step is followed by a similar coarse-graining in $x$-direction and these directions are alternated. The same algorithm can be used if the indices of the initial tensors are permuted accordingly after each coarse-graining step. There are four different choices to exchange the $x$ and $y$ directions, which are also shown in \cref{Appfig:rotExplanation}:

\beqnn{
K_{xyx'y'} 
&\leftrightarrow
K_{yxy'x'}&\mathrm{\ \ \ (x \leftrightarrow y),}\\
K_{xyx'y'} 
&\leftrightarrow
K_{y'x'yx}&\mathrm{\ \ \ (x \leftrightarrow y'),}\\
K_{xyx'y'} 
&\rightarrow
K_{yx'y'x}&\mathrm{\ \ \ (\circlearrowleft),}\\
K_{xyx'y'}
&\rightarrow
K_{y'xyx'}&\mathrm{\ \ \ (\circlearrowright)}.
}

\begin{figure}[h!]
\begin{center}
 \includegraphics[width=8cm, angle=0]{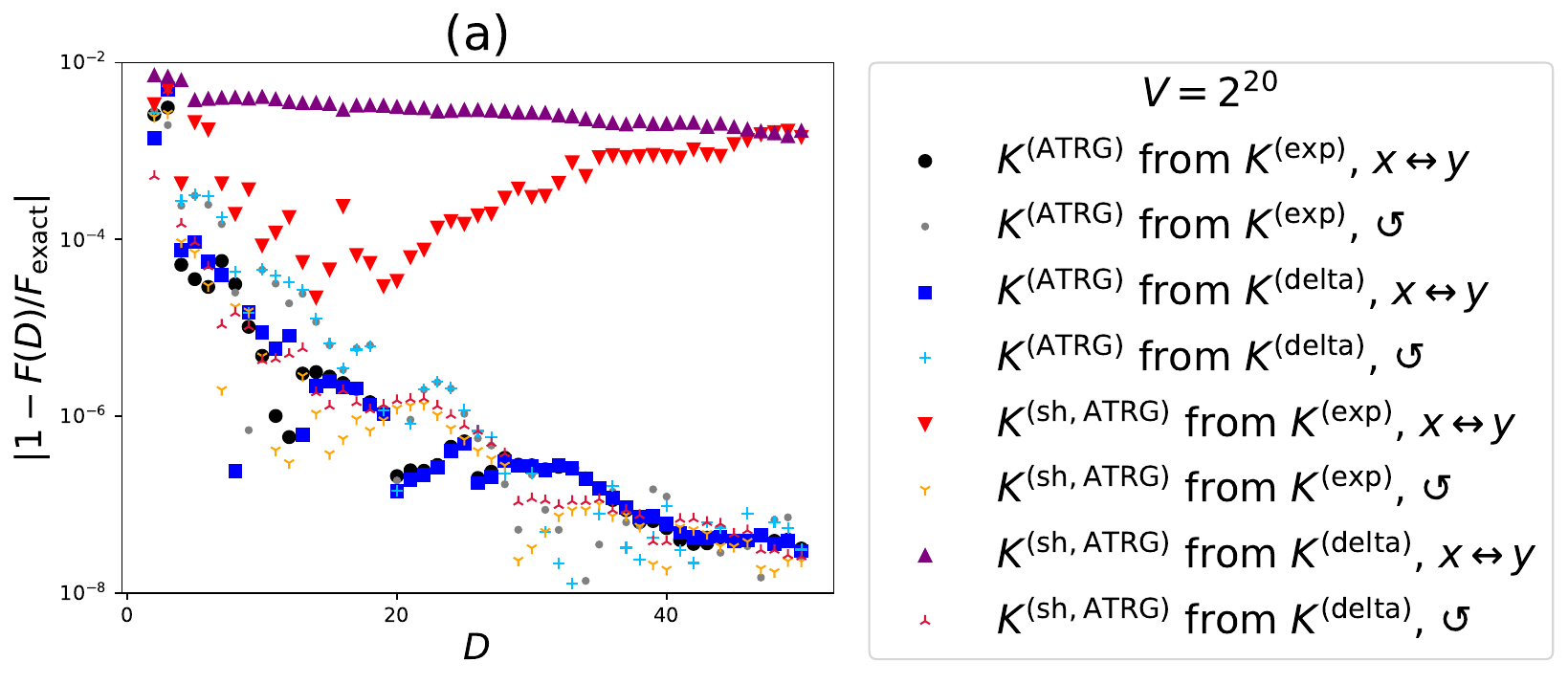}
 \includegraphics[width=8cm, angle=0]{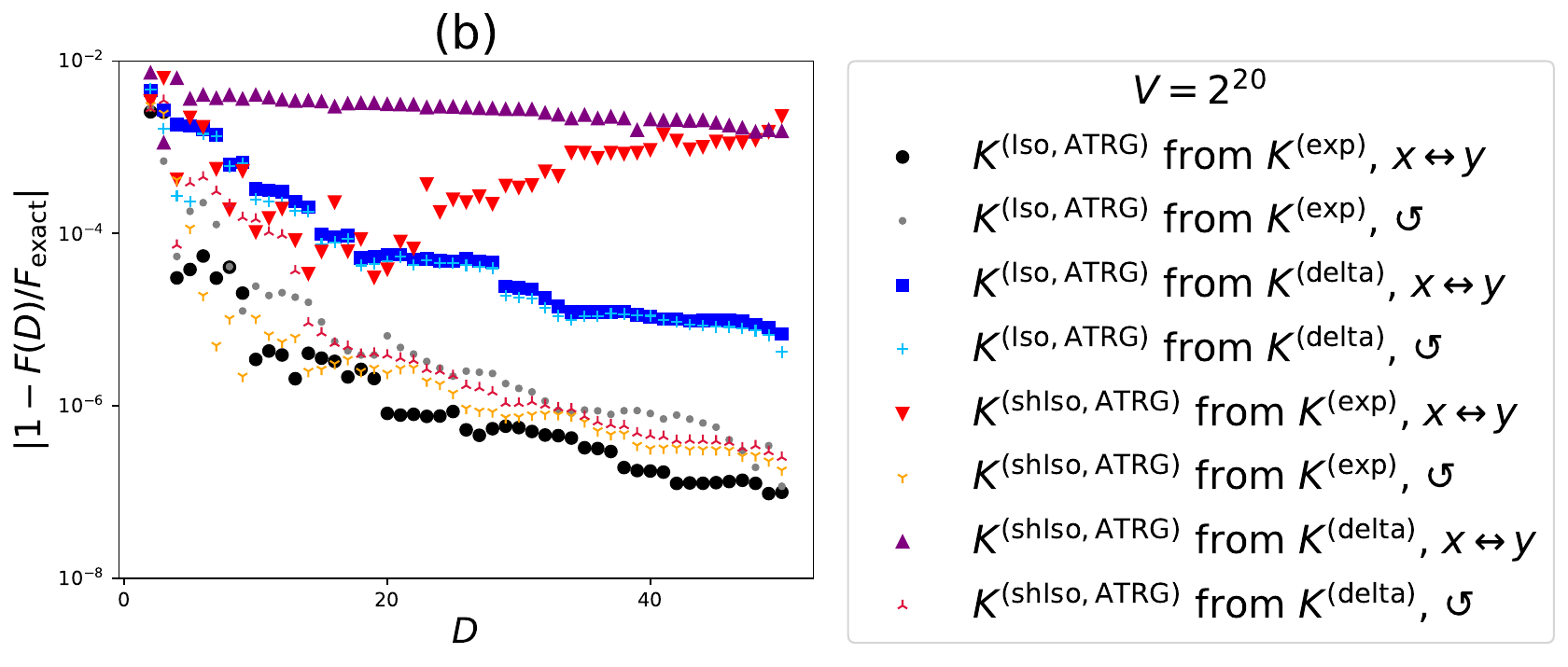}
 \includegraphics[width=8cm, angle=0]{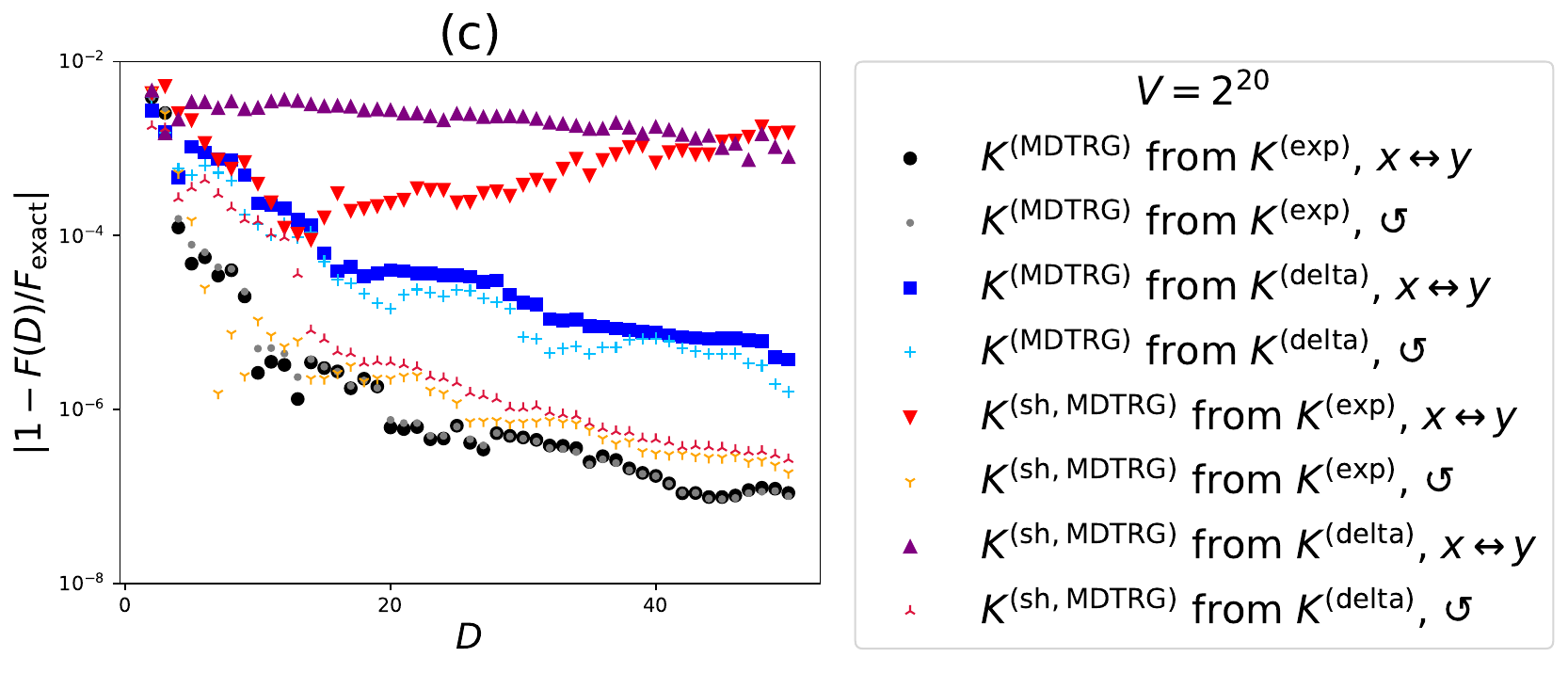}
 \includegraphics[width=8cm, angle=0]{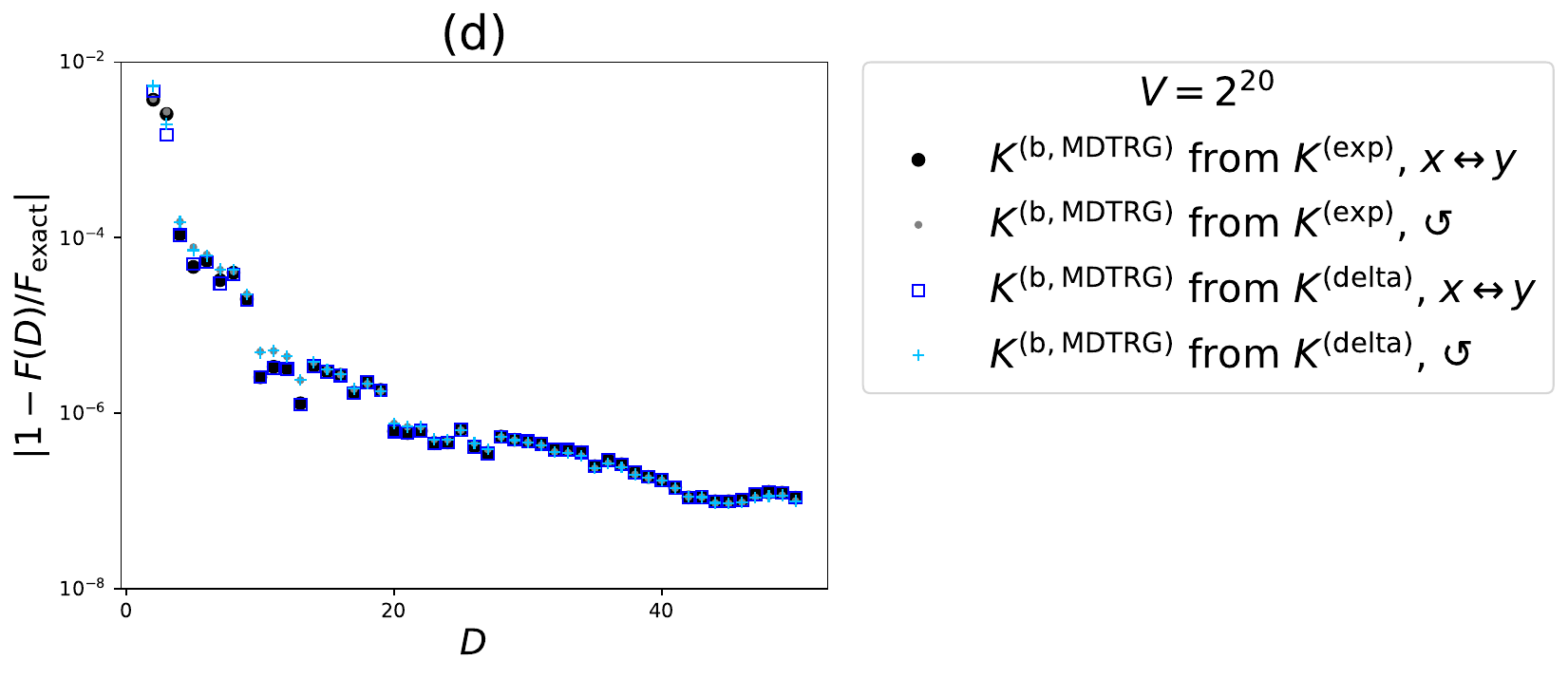}
 \caption{
 Dependence of TRG algorithms on the type of exchange between $x-$ and $y-$directions between coarse-graining steps. Shown is the relative error of the free energy for the critical Ising model in two-dimensions for different bond dimensions $D$ and for a symmetric initial tensor $K^\mathrm{(exp)}$ and a non-symmetric initial tensor $K^\mathrm{(delta)}$. Comparison between different ATRG and MDTRG algorithms as introduced in \cref{app:ATRG_and_Triad}.
}
\label{Appfig:rot}
\end{center}
\end{figure}

\begin{table}[t!]
 \centering
\begin{tabular}{l l l l } 
 \hline \hline
   & $xy-$swap dep. & Trunc. & $K$ dep. \\ 
\hline
ATRG~\cite{ATRG}& $-$ & \textit{sqz} &$--$ \\
Iso-ATRG~\cite{ATRG}&$+$  & \textit{iso} &$++$ \\
sh-ATRG & $++$ & \textit{sqz} &$-$ \\
sh-Iso-ATRG &$++$  & \textit{iso*} &$-$\\
\hline
MDTRG~\cite{MDTRG} & $-$ & \textit{iso} &$++$\\
sh-MDTRG &$++$  & \textit{iso*} &$-$\\
b-MDTRG &$--$  & \textit{sqz} &$--$\\
\hline \hline  
 \end{tabular}
 \caption{
 Properties of different ATRG and MDTRG methods. 2nd column: dependence on the type of exchange between $x-$ and $y-$direction between coarse-graining steps; $++$/$+$/$-$/$--$ stands for very strong/noticeable/slight but not significant/nearly no dependence; 3rd column: truncation method; \textit{iso} stands for isometries which are used to create the coarse-grained indices; \textit{iso*} means that isometries are used for intermediate approximate contractions, but they do not create the new indices of the coarse-grained tensors directly; \textit{sqz} denotes all other methods, so either the squeezers from boundary TRG~\cite{boundaryHOTRG} (see main text and \cref{app:b_TRG}), or a simple contraction and singular value decomposition. 4th column: dependence on the initial tensors; $--$ stands for no dependence, $-$ for a slight but not significant dependence, $++$ for strong dependence;
 } \label{tab:ATRGMDTRG_rot}
\end{table}

We test the dependence of the TRG variants on the type of $xy$-exchange, see \cref{Appfig:rot}. 
We only show data for $x\leftrightarrow y$ and $\circlearrowleft$ because the results for $x\leftrightarrow y$ and $\circlearrowleft$ coincide with $x\leftrightarrow y'$ and $\circlearrowright$, respectively. 
We summarize our findings in \cref{tab:ATRGMDTRG_rot}.
The main observations from the numerical benchmarks are:
\begin{enumerate}
\item The shifted methods with a flip $x\leftrightarrow y$ do not converge to the correct results when the bond dimension is increased, and the errors remain large or even increase with the bond dimension (red and purple triangles in \cref{Appfig:rot}).
\item Non-shifted methods have a similar or better accuracy when a flip $x\leftrightarrow y$ is applied. The results are in particular better for the isometric ATRG (black and gray dots in \cref{Appfig:rot}(c)).
\item The boundary TRG methods do not significantly depend on the type of exchange, $x\leftrightarrow y$ or rotation $\circlearrowleft$.
\item Overall, the different types of exchange (rotation $\circlearrowleft$ or flip $x\leftrightarrow y$) lead to different accuracies, depending on the details of the coarse-graining algorithm. Therefore, the exchange type should be chosen accordingly.
\end{enumerate}

When implementing a TRG algorithm, one has to carefully keep track of the index order and conventions. For example, we identified $f$ in \cref{eq:shATRG_J} as $X$ in \cref{eq:shATRG_K}, and $f'$ as $X'$. If one would instead set $X'=f$ and $X=f'$, it would correspond to an exchange between the $xy$ flip and a rotation. These conventions should be explicitly checked when comparing flips and rotations between algorithms and implementations.

The observations can be understood from the interplay of the last step in obtaining a coarse grained tensor, the exchange of indices, and the initial tensor decomposition in a TRG algorithm. For example, the coarse grained tensor $K^\mathrm{(ATRG)}$ in the ATRG algorithm is obtained by a contraction of two tensors $M$ and $L$ in \cref{eq:Katrg} and \cref{fig:ATRG_schpic} from (b) to (c). A flip $x \leftrightarrow y$ ($x \leftrightarrow y'$) exchanges two indices of $M$ and two indices of $L$, but does not move only one index to the other tensor. In the next coarse-graining iteration, the tensor $K^\mathrm{(ATRG)}$ is initially split into $E$ and $F$, which are exactly $M$ and $L$ ($L$ and $M$) respectively. Therefore, this SVD does not introduce a further truncation. This is not be the case if a rotation of the indices is used. Similarly, the initial splitting of $K^\mathrm{(sh,ATRG)}$ into $E$ and $F$ for the shifted ATRG reconstructs the tensors $M$ and $L$ ($L$ and $M$) respectively if the exchange type $\circlearrowleft$ ($\circlearrowright$) is used. This can be seen from \cref{eq:shATRG_J,eq:shATRG_K} or \cref{fig:ATRG_schpic}(b) to (c). The same arguments hold for the MDTRG algorithms.

The optimal choice for the index exchange can also be understood if the triad representation is used everywhere instead of coarse-graining to a square lattice~\cite{TriadRG,MDTRG,RandTRG}. For example, the tensor $K^\mathrm{(ATRG)}$ does not need to be constructed explicitly as a contraction between $M$ and $L$. Instead, these two tensors can be used in the next coarse graining step. In this formulation, the natural index exchange order is more apparent.

We benchmarked the two-dimensional Ising model at the critical temperature here. Since we discover a significant dependence on the type of $xy$-swapping for some of the methods, we suggest to check this behavior for other models as well to find the optimal choice. 
This is particularly true since we found specific cases for the flip-index exchange $x\leftrightarrow y$ with a systematic accumulation of errors, which led to a decreased accuracy when the bond dimension is increased. 
Similarly, the type of index permutation after each coarse-graining step can be important for other TRG methods and in higher dimensions, where the number of variants becomes even larger.

\bibliography{bibliography}

\end{document}